%% file: lightscalar.tex
\documentclass[preprint,aps,prd]{revtex4-1}

\usepackage{amsmath}
\usepackage{graphicx}
\usepackage{hyperref}
\usepackage[T1]{fontenc}
\usepackage{slashed}
\usepackage{bbold}

\newcommand{\be}{\begin{equation}}
\newcommand{\ee}{\end{equation}}

\newcommand{\ckm}{V_{\mathrm{CKM}}}
\DeclareMathOperator{\Tr}{Tr}

\usepackage[normalem]{ulem}

\begin{document}

\title{Flavor-specific scalar mediators}

\author{Brian Batell}
\email{batell@pitt.edu}
\author{Ayres Freitas}
\email{afreitas@pitt.edu}
\author{Ahmed Ismail}
\email{aismail@pitt.edu}
\author{David McKeen}
\email{dmckeen@pitt.edu}
\affiliation{Pittsburgh Particle Physics, Astrophysics, and Cosmology Center,\\Department of Physics and Astronomy, University of Pittsburgh, Pittsburgh, USA}

\begin{abstract}
New singlet scalar bosons have broad phenomenological utility and feature prominently in many extensions of the Standard Model. Such scalars are often taken to have Higgs-like couplings to SM fermions in order to evade stringent flavor bounds, e.g. by assuming Minimal Flavor Violation (MFV), which leads to a rather characteristic phenomenology. 
Here we describe an alternative approach, based on an effective field theory framework for a new scalar that dominantly couples to one specific SM fermion mass eigenstate. A simple flavor hypothesis ensures adequate suppression of new flavor changing neutral currents. We consider radiatively generated flavor changing neutral currents and scalar potential terms in such theories, demonstrating that they are often suppressed by small Yukawa couplings, and also describe the role of $CP$ symmetry. We further demonstrate that such scalars can have masses that are significantly below the electroweak scale while still being natural, provided they are sufficiently weakly coupled to ordinary matter. In comparison to other flavor scenarios, our framework is rather versatile since a single (or a few) desired scalar couplings may be investigated in isolation. We illustrate this by discussing in detail the examples of an up-specific scalar mediator to dark matter and a muon-specific scalar that may address the $\sim 3 \sigma$ muon anomalous magnetic moment discrepancy.
\end{abstract}

\maketitle

\section{Introduction}
\label{sec:intro}

Despite its many successes, the Standard Model (SM) is widely suspected of being incomplete. Along with the empirical mysteries of dark matter, the matter-antimatter asymmetry, and neutrino masses, the naturalness of the Higgs boson is often cited as a motivation for new physics. In the SM the Higgs is described as a fundamental scalar field, and experimental studies of its properties at the LHC are so far  consistent with this description. However, as is well-known, fundamental scalar masses are quadratically sensitive to new ultraviolet (UV) physics scales, suggesting that new physics should appear near the electroweak scale. While this expectation has not yet been borne out by experiment (hence the naturalness problem), such reasoning has had clear successes in the past, e.g.,~ the charged/neutral pion mass splitting in QCD.

Against the backdrop of exploration at the energy frontier, recent years have seen renewed interest in the possibility of light hidden sectors containing new SM gauge singlet states with masses well below the weak scale. 
In particular, new light scalar particles play a prominent role in many of these scenarios. To mention a few examples, 
light scalars could help resolve outstanding theoretical issues, such as the strong CP problem~\cite{Peccei:1977np,Peccei:1977hh,Weinberg:1977ma,Wilczek:1977pj} (a naturalness question itself), 
be responsible for hidden sector mass generation (via a ``dark'' Higgs mechanism), 
mediate interactions between the SM and dark matter (DM) or even comprise the DM \cite{Silveira:1985rk,McDonald:1993ex,Burgess:2000yq,Boehm:2003hm,Borodatchenkova:2005ct,Fayet:2006sp}, 
or provide an explanation of various experimental anomalies (e.g., the muon anomalous magnetic moment discrepancy~\cite{Gninenko:2001hx,Fayet:2007ua,Pospelov:2008zw,Davoudiasl:2012ig}). 
In particular, light scalars have been explored in multiple contexts and comprise an interesting class of phenomenologically motivated
theories \cite{TuckerSmith:2010ra,Batell:2011qq,Schmidt-Hoberg:2013hba,Clarke:2013aya,Chen:2015vqy,Batell:2016ove}. 
Of course, any additional fundamental scalar would suffer from the same naturalness
problem as the Higgs, and for scalars lighter than the electroweak scale, the
required tuning is potentially even more severe. While light scalars have some advantages
over their spin-1 counterparts, such as the lack of a need to cancel gauge
anomalies which can lead to stringent bounds~\cite{Preskill:1990fr,Batra:2005rh,Dror:2017ehi,Ismail:2017ulg,Dror:2017nsg,Ismail:2017fgq}, naturalness suggests that they
should not appear in isolation unless they are sufficiently weakly coupled. In this paper, we seek to estimate the implications of naturalness for a generic light scalar coupled to SM fermions.

Along with technical naturalness considerations, a basic issue that arises in scenarios with light scalars pertains to the structure of their couplings to SM particles. Often one or a few couplings are postulated for some desired phenomenological purpose and then studied in isolation (see for example Refs.~\cite{TuckerSmith:2010ra,Omura:2015nja,Carlson:2015poa}) while other allowed couplings are neglected. 
Can such a starting point be justified in an effective field theory approach, and can it be consistent with a host of experimental bounds from flavor physics? Perhaps the simplest way to avoid new flavor changing neutral currents (FCNCs) is to impose a symmetry principle such as Minimal Flavor Violation (MFV)~\cite{DAmbrosio:2002vsn}. Such a scenario, while certainly well motivated, implies that the scalar preferentially couples to the third generation fermions and does not offer the flexibility needed for all phenomenological applications. 
Several extensions of MFV have been considered, often in the context of heavy new physics which couples only to the third generation of the SM. Here we pursue an alternative approach to MFV by considering
couplings that are specific to \emph{any} one SM fermion. By treating interactions with non-trivial flavor structure as spurions, we will see that a single new coupling can often naturally dominate the phenomenology of a theory with an appropriate flavor symmetry principle. Our goal in this paper is not to propose a mechanism for generating such a single-fermion flavor pattern from a dynamical origin or a fundamental symmetry, but instead to study constraints from self-consistency and elucidate the phenomenological consequences of such a scenario.

Our results have implications for any new light scalar, which would be badly tuned without satisfying the guidelines we present. We show two examples, demonstrating the applicability of our construction to a scalar that couples to muons to resolve the discrepancy between the observed and predicted anomalous magnetic moment of the muon, as well as 
a scalar that couples preferentially to up quarks and mediates interactions with dark matter (a realization of ``leptophobic'' dark matter). 
Often, the range of natural couplings is only now being probed experimentally.

The remainder of this paper is organized as follows. In the next section, we study the impact of a new scalar with a single coupling to a SM fermion. From symmetry arguments, we estimate the sizes of the scalar's couplings to the SM as well as its potential. In Section \ref{sec:applications}, we apply our considerations of naturalness to particular models of light scalars, comparing the natural regions of parameter space with the reach of current and future experiments. Section \ref{sec:concl}
 contains our conclusions.

\section{Effective field theory of a flavor-specific scalar}
\label{sec:symmetries}

In this section we present an effective field theory framework describing a new light scalar particle $S$ with flavor-specific couplings. We use the term ``flavor-specific'' to mean that the scalar dominantly couples to a particular SM fermion mass eigenstate. We will describe how a simple flavor hypothesis in the effective field theory ensures the adequate suppression of new FCNCs. We also investigate the natural sizes of radiatively generated couplings and scalar potential interactions, which will lead to a naturalness criterion in the physical scalar mass - coupling parameter space. Following the presentation of the EFT framework in this section, we will present two phenomenological  applications in Section~\ref{sec:applications}.

We begin by reviewing the application of flavor symmetries to theories of new physics, using the MFV hypothesis as a starting point.
We write the SM gauge and Yukawa interactions of the quarks as
\be
\label{eq:smlag}
\mathcal{L}_\mathrm{SM} = i \bar{Q}_L \slashed{D} Q_L + i \bar{U}_R \slashed{D} U_R + i \bar{D}_R \slashed{D} D_R - \left( \bar{Q}_L Y_u U_R H_c + \bar{Q}_L Y_d D_R H + \mathrm{h.c.} \right),
\ee
where $Q_L = \begin{pmatrix} U_L \\ D_L \end{pmatrix}$ and $H$ is the Higgs doublet with $H_c=i\sigma^2H^\ast$. For conciseness, we will focus on the quark sector, pointing out differences from the lepton case as necessary. Throughout, we use 4-component notation with implied projection operators, e.g.~the right-handed up quark is $U_R \equiv P_R u$, where $u$ is the usual up quark. The Yukawa interactions break the full 
$U(3)_Q \times U(3)_U \times U(3)_D$ global flavor 
symmetry to $U(1)_{B}$ baryon number.\footnote{Of course, hypercharge is also conserved. Including a global $U(1)_H$ factor for the Higgs, the full breaking pattern is $U(3)_Q \times U(3)_U \times U(3)_D \times U(1)_H \rightarrow  U(1)_{B} \times U(1)_Y$.}
In the presence of new physics, MFV postulates that the SM Yukawas are the \emph{only} couplings which break the flavor symmetry~\cite{DAmbrosio:2002vsn}. To estimate the size of flavor-violating effects, the flavor symmetry may be formally restored by treating the Yukawa couplings as bifundamentals under $SU(3)^3$, namely $Y_u \sim (3, \bar{3}, 1)$ and $Y_d \sim (3, 1, \bar{3})$, and requiring that new physics operators are flavor singlets. 

In anticipation of our flavor-specific flavor hypothesis, it will be instructive to examine the symmetry breaking of $Y_u$ and $Y_d$ in isolation. Consider first the case $Y_u \neq 0$ and $Y_d = 0$. In this case, the $U(3)_D$ symmetry is unbroken, while general $Y_u$ leads to the breaking pattern
\begin{equation}
\label{eq:Yu-break}
U(3)_Q \times U(3)_U \rightarrow U(1)_u \times U(1)_c \times U(1)_t    ~~~~~(Y_u \neq 0, Y_d = 0).
\end{equation}
That is, in the limit $Y_d = 0$, there is a $U(1)^3$ quark flavor symmetry that acts on the physical up-type quark mass eigenstates. Since $U(3)_D$ symmetry is unbroken, it is possible to re-phase the right-handed down quarks in order to identify an unbroken $U(1)^3$ baryon flavor symmetry which re-phases the three generations of baryons.
Similarly, in the case $Y_u = 0$ and $Y_d \neq 0$, the $U(3)_U$ symmetry is preserved, while general $Y_d$ leads to the breaking pattern
\begin{equation}
\label{eq:Yd-break}
U(3)_Q \times U(3)_D \rightarrow U(1)_d \times U(1)_s \times U(1)_b   ~~~~~(Y_u = 0, Y_d \neq 0),
\end{equation}
i.e., there is a $U(1)^3$ quark flavor symmetry that acts on the physical down-type quark mass eigenstates, which can be extended to a $U(1)^3$ baryon flavor symmetry. Now, consider again the case of both $Y_u$ and $Y_d$ non-vanishing (the case of the SM). Because the CKM matrix is nontrivial, the remnant $U(1)^3$ quark flavor symmetries preserved by $Y_u$ (in Eq.~(\ref{eq:Yu-break})) and $Y_d$ (in Eq.~(\ref{eq:Yd-break})) are different, and only the full $U(1)_B$ baryon number symmetry remains. 

We now add a real SM singlet scalar $S$ which can interact with the quarks through dimension-five operators. Broadly speaking, such couplings can either take place through $\partial S$ or $S$ itself, viz.
\be
\label{eq:bsmlag}
\begin{aligned}
\mathcal{L}_S = \frac{1}{2} \partial_\mu S \partial^\mu S - \frac{1}{2} m_S^2 S^2 &- \biggl( \frac{c_S}{M} S \bar{Q}_L U_R H_c + \mathrm{h.c.} \biggr)
+ \frac{d_S}{M} \partial_\mu S \bar{U}_R \gamma^\mu U_R \\
&+\frac{d'_S}{M}  \biggl( i S \bar{U}_R \slashed{D} U_R + \mathrm{h.c.} \biggr).
\end{aligned}
\ee
where $c_S$ is a complex $3\times3$ matrix and $d^{(\prime)}_S={d^{(\prime)}_S}^\dagger$ are Hermitian $3\times3$ matrices.
Here we have only written three possible couplings, though interactions analogous to the third term in Eq.~\eqref{eq:bsmlag} but with the down-type quarks, as well as interactions analogous to the fourth and fifth terms in Eq.~\eqref{eq:bsmlag} but with left-handed quarks or right-handed down-type quarks are also possible. Including these, for $N$ flavors, there are $2 N^2$ possible complex couplings of the $c_S$ type and $6 N^2$ real couplings of the $d^{(\prime)}_S$ type in the above. The couplings $c_S$, $d_S$, and $d'_S$ carry flavor indices, like the SM Yukawas, and any flavor hypothesis such as MFV restricts their form. If $S$ is a flavor singlet, the couplings in Eq.~\eqref{eq:bsmlag} have the flavor structure
\begin{align}
c_S &\sim (3, \bar{3}, 1), \nonumber \\
d_S, d'_S &\sim (1, 1, 1) \oplus (1, 8, 1).
\end{align}
For instance, under MFV, $c_S = c_1 Y_u + \dots $, while $d_S = d_1 \mathbb{1} + d_2 Y_u^\dag Y_u + \dots$.

The three types of operators represented by the interaction terms in Eq.~\eqref{eq:bsmlag} can be shown to be related to each other through appropriate field redefinitions. 
Starting from a theory with $d_S, d'_S \neq 0$, 
we can perform the transformation
\be
U_R \rightarrow U_R - (d'_S - i d_S) S U_R / M,
\label{eq:Uredef}
\ee
which removes the $d^{(\prime)}_S$ terms at the expense of inducing a $c_S$ term with strength $c_S = -Y_u(d'_S - i d_S)$
plus an additional dimension-six higher derivative operator.  Note that the strength of the induced $S\bar{Q}UH$ coupling is proportional to the Yukawa coupling and is thus suppressed for light quarks (i.e., the induced $c_S$ has an MFV-like flavor structure if $d_S$ and $d_S^\prime$ are proportional to the identity).
Through analogous field redefinitions for the left-handed quarks and right-handed down-type quarks, we may eliminate all of the $d^{(\prime)}_S$-type terms of Eq.~\eqref{eq:bsmlag}.

Here we wish to consider flavor-specific flavor structures which are not found under the MFV hypothesis. In particular, we will be interested in the possibility that the dominant couplings of $S$ are to the first or second generation fermions in the zero momentum limit. 
We find it convenient to work with an operator basis where the $d^{(\prime)}_S$-type terms are eliminated through the field redefinitions described above. The $c_S$-type terms contain the full information of the couplings of $S$ to quarks, with the only considerations for their structure coming from the flavor-specific flavor hypothesis which we describe in more detail below. Once we make such a hypothesis, we are no longer working with the most general version of Eq.~\eqref{eq:bsmlag}. Below, we motivate and describe the particular $c_S$ flavor structure with which we are concerned.

Note that, in the case of a single flavor-specific coupling, by inverting the field redefinition of Eq.~(\ref{eq:Uredef}) to generate $d_S$ and $d_S^\prime$ operators, we see that the real part of $c_S$ breaks the shift symmetry of $S$, while the imaginary part of $c_S$ seemingly preserves the shift symmetry since the leading operator to which it leads involves $\partial_\mu S$. However, this shift symmetry is broken by a dimension-six operator that is induced by this field redefinition,
\be
\frac12\left|\frac{c_S}{Y_u}\right|^2\left(\frac{S}{M}\right)^2\biggl( i \bar{U}_R \slashed{D} U_R + \mathrm{h.c.} \biggr),
\ee
although a purely imaginary $c_S$ preserves a parity symmetry under which $S\to -S$. One of our primary goals will be to understand the natural size and physical consequences of the induced scalar potential.

Besides MFV, there are other flavor symmetry principles that can lead to viable flavor phenomenology. 
One example is next to minimal flavor violation (NMFV), which assumes that new physics couples dominantly to the third generation~\cite{Agashe:2005hk}. This case is distinct from the MFV hypothesis; while the new physics breaks the $U(3)^3$ quark flavor symmetry in a way that is not proportional to the SM Yukawas, it preserves a $U(2)^3$ symmetry that is only broken by the SM. 
In general, the chiral symmetry broken by new physics need not be aligned with that of any of the usual SM Yukawas. However, assuming a limited set of flavor-breaking spurions in NMFV ensures that flavor mixing effects between the third and the first two generations is not parametrically larger than in the SM \cite{Barbieri:2012uh,Barbieri:2012bh}, i.e.\ the new physics and Yukawa interactions are quasi-aligned up to extra mixing contributions that are not parametrically larger than the CKM mixing angles.

Here we wish to explore instead the hypothesis that the new physics coupling $c_S$ involves only a single fermion, in the mass eigenstate basis. This hypothesis is a more restrictive case of the alignment hypothesis. Alignment requires that $c_S$ and the Yukawa interactions are simultaneously diagonalizable in a single basis. We will further assume that new physics  couples to only one fermion mass eigenstate. This implies that the spurion $c_S$ breaks the $U(3)^3$ flavor symmetry in a specific way that is determined based the particular fermion that  couples to $S$.
To see this, consider the flavor symmetry breaking induced by a scalar that couples specifically to the up quark in the mass basis, $c_S \propto {\rm diag}(1,0,0)$. In spurion language, this assumption is equivalent to assuming that $c_S$ breaks the flavor symmetry as follows,
\begin{equation}
\label{eq:cS-break}
U(3)_Q \times U(3)_U \rightarrow U(1)_u \times U(2)_{ctL} \times U(2)_{ctR}.
\end{equation}
In particular, it is crucial that the $U(1)_u$ factor in Eq.~\eqref{eq:cS-break} is the same as the one left unbroken by $Y_u$ in Eq.~\eqref{eq:Yu-break}. 

The alignment hypothesis is possibly mysterious from an bottom up perspective, and raises the specter of significant fine tuning. 
We will not endeavor to construct a flavor model or mechanism which naturally achieves alignment in this work, although we note that there are some promising model building approaches in the literature~\cite{Knapen:2015hia,Altmannshofer:2017uvs}.

It is worthwhile to compare the flavor-specific hypothesis to MFV. In MFV, the basic assumption is that the Yukawa couplings are the only spurions that break flavor, and therefore new FCNCs are generally SM-like. In our flavor-specific scenario, in addition to the Yukawa couplings, we are adding a new flavor-breaking spurion $c_S$, and assume that it is aligned with the Yukawa couplings according to 
Eqs.~(\ref{eq:Yu-break},\ref{eq:Yd-break},\ref{eq:cS-break}). 
In this sense, the flavor-specific hypothesis we are exploring rests on stronger assumptions about how the $U(3)^3$ flavor symmetry is broken in the UV.

In this framework, the couplings which violate the flavor and scalar shift symmetries are the SM Yukawas, $c_S$ and $m_S$. Assuming that these are the leading symmetry-violating effects, we may estimate the size of any operator in the effective field theory through spurion analysis. In the following, we will describe the sizes of the operators $S^n$ and $S \bar{Q}_L D_R H$, respectively. First, however, we consider corrections to each of our original couplings themselves.

\subsection{Naturalness of leading couplings}
\label{subsec:EFTnaturalness}
Here, we wish to use symmetry arguments to estimate the sizes of corrections to the SM Yukawas, $c_S$, and $m_S$, assuming they are the only leading interactions
\be
\mathcal{L} \supset - \frac{1}{2} m_S^2 S^2 - \left( \bar{Q}_L Y_u U_R H_c + \frac{c_S}{M} S \bar{Q}_L U_R H_c + \mathrm{h.c.} \right).
\ee
We first observe that both the couplings $c_S$ and $Y_u$ break the up-type quark chiral symmetry, while $c_S$ additionally breaks the $S$ shift symmetry. $Y_d$ breaks the down-type quark chiral symmetry. The $S$ mass breaks the $S$ shift symmetry only.

By treating $c_S$ and $Y_u$ as spurions, it follows immediately that they are technically natural. When $S$ acquires a vacuum expectation value (vev) $v_S$ so that $c_S$ and $Y_u$ are no longer distinguished by their $S$ shift symmetry properties, then $c_S$ immediately leads to the induced up Yukawa
\be
\delta Y_u = \frac{c_S v_S}{M}.
\ee
We will return to this constraint in Section~\ref{sec:potential}, after estimating the natural size of $v_S$.

\begin{figure}[tbp]
\centering
\includegraphics[width=\textwidth]{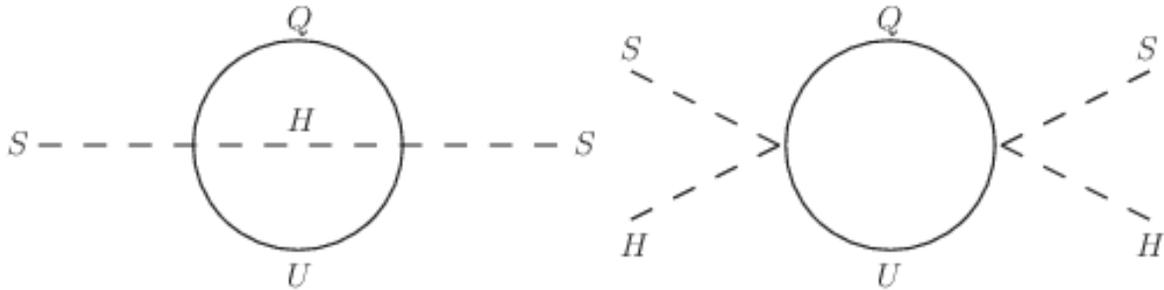}
\caption{Diagrams correcting the scalar mass in the effective theory of $\mathcal{L}_S$ in Eq.~\eqref{eq:bsmlag}.}
\label{fig:smasscorreft}
\end{figure}

Finally, as $S$ is a scalar, its mass is \emph{not} natural, and suffers from the usual hierarchy problem. If we assume that new physics comes in at the scale $M$ to regulate corrections to the $S$ mass, however, we may still obtain useful naturalness constraints on the interactions in $\mathcal{L}_S$. In particular, the $S$ mass is corrected by the diagrams of Figure~\ref{fig:smasscorreft}. The two-loop diagram leads to a mass shift of order
\be
\label{eq:s2bound}
\delta m_S^2 \sim \frac{\Tr c_S^\dagger c_S}{(16\pi^2)^2} M^2.
\ee
Requiring that this be less than the $S$ mass squared itself yields the naturalness criterion 
\be
\label{eq:csboundfroms2}
(c_S)^{ij} \lesssim (16\pi^2) \frac{m_S}{M} \approx (3 \times 10^{-3}) \left( \frac{m_S}{0.1~\mathrm{GeV}} \right) \left( \frac{5~\mathrm{TeV}}{M} \right)
\ee
on the elements of $c_S$.\footnote{A similar constraint could be placed using the Higgs mass correction, but it would be weaker for $m_S < m_h$.}

The Higgs portal operator $S^2 H^2$, which is generated from the one-loop diagram in Figure~\ref{fig:smasscorreft}, also leads to an $S$ mass correction after electroweak symmetry breaking
\be
\delta m_S^2 \sim \frac{\Tr c_S^\dagger c_S}{32\pi^2} v^2
\ee
leading to the bound
\be
\label{eq:s2h2bound}
(c_S)^{ij} \lesssim (4 \pi \sqrt{2}) \frac{m_S}{v} \approx (7 \times 10^{-3}) \left( \frac{m_S}{0.1~\mathrm{GeV}} \right).
\ee
The relative importance of these two constraints depends on the size of the cutoff scale $M$. For $M$ above (below) a few TeV, the bound in Eq.~\eqref{eq:csboundfroms2} (Eq.~\eqref{eq:s2h2bound}) is stronger.

\subsection{Scalar potential}
\label{sec:potential}

We have estimated the corrections to the operators in Eq.~\eqref{eq:bsmlag} in the previous section. In general, additional operators will also be generated. Here we estimate the size of radiatively generated $S^n$ terms for arbitrary $n$, assuming that they are zero at tree level.

The only interaction involving the new scalar is the $c_S$ coupling, which involves one $S$ field. Consequently, the radiative generation of $S^n$ requires $n$ insertions of $c_S$. In addition, since $S^n$ preserves the chiral quark symmetries, if $n$ is odd we must have at least one quark Yukawa as well (or an $S$ vev). Therefore, the natural sizes of the $S^n$ operators are
\be
\begin{split}
\label{eq:sncorr}
\delta_{S^{2k}} &\sim \frac{\Tr (c_S^\dagger c_S)^k}{(16\pi^2)^{k + 1}} M^{4 - 2k},\ k = 1, 2, \ldots\\
\delta_{S^{2k + 1}} &\sim \frac{\Tr (c_S^\dagger c_S)^k c_S^\dagger Y_u}{(16\pi^2)^{k + 2}} M^{4 - (2k + 1)},\ k = 0, 1, \ldots.
\end{split}
\ee
Note that there are multiple possible flavor contractions in the above.

As before, we also get a contribution to $S^n$ from the operators $S^n H^{2m}$. The relevant diagrams may be constructed by cutting $m$ Higgs propagators to break loops, e.g. as in the diagrams of Figure~\ref{fig:smasscorreft}. Each cut gives two extra Higgs vevs which replace the cutoff scale $M$, and eliminates one loop, so we expect the correction $\delta_{S^n}$ from the operator $S^n H^{2m}$ to be related to the correction in Eq.~\eqref{eq:sncorr} by the factor $\left( \frac{8 \pi^2 v^2}{M^2} \right)^m$. For $M$ larger than a few TeV, this factor is a suppression, while for $M$ smaller than a few TeV it is an enhancement.

The radiatively generated $S^n (H^{2m})$ terms lead to a scalar potential which we should minimize to obtain the $S$ and $H$ vevs. Assuming large $M$, we neglect operators with $m > 0$ and minimize $V(S)$ alone. If the potential terms involving both $S$ and $H$ are small relative to $V(H)$ after inserting the $S$ vev, they will not significantly affect the minimization of the usual Higgs potential. We remark in particular that the $S-H$ mixing is small for large cutoff scales. In particular, the radiatively generated $S H^2$ term induces a mixing that is roughly $\frac{\Tr c_S^\dagger Y_u}{(16\sqrt{2}\pi^2)} v M S H$. If the coupling $c_S$ satisfies the naturalness bound of Eq.~\eqref{eq:csboundfroms2}, then the mixing angle in the scalar sector is at most
\be
\sin \theta_{SH} \lesssim \frac{Y_u^i v m_S}{\sqrt{2} m_h^2}
\ee
for coupling to a single up-type quark $u^i$, which is small for light $S$ and especially for a scalar that couples only to a first- or second-generation quark. 

For $c_S$ satisfying the naturalness bound in Eq.~\eqref{eq:csboundfroms2} and a significant hierarchy $M \gg m_S$, the linear and quadratic terms dominate the $S$ potential. This is not surprising since higher dimension operators are suppressed by factors of the small $c_S$, as well as additional loops. Given the tadpole term $\delta_S S$, which can be estimated using Eq.~\eqref{eq:sncorr}, the resulting scalar vev is
\be
\label{eq:svev}
v_S \approx -\frac{\delta_S}{m_S^2} \sim \frac{\Tr c_S^\dagger Y_u}{(16\pi^2)^2} \left( \frac{M}{m_S} \right)^2 M.
\ee

The scalar vev induces corrections to the quark masses. From the dimension-five operator involving $c_S$, we have the mass correction
\be
\label{eq:mqcorr}
\delta m_{u^i} = \frac{c_S^{ii} v_S v}{\sqrt{2} M}.
\ee
For large $M$, inserting the vev of Eq.~\eqref{eq:svev} and requiring that $\delta m_{u^i} \lesssim m_{u^i}$ yields
an identical bound to Eq.~\eqref{eq:csboundfroms2}.

In principle, $v_S$ also leads to a correction to $m_S$ from operators of the form $S^n$ with $n > 2$, which in turn limits $c_S$. However, if these operators are only radiatively generated, these effects are minor since the linear and quadratic terms dominate the $S$ potential. For a scalar coupling only to the up-type quark $u^i$, the corrections to the $S$ mass from the $S^n$ operators go as
\be
\begin{split}
\delta^{(2k)}_{m_S^2} \sim \delta_{S^{2k}} v_S^{2k - 2} \lesssim m_S^2 &\to (c_S)^{ii} \lesssim \left(16 \pi^2 \right)^{\frac{5k -3}{4k -2}} (Y_u^i)^{-\frac{2k - 2}{4k - 2}} \left( \frac{m_S}{M} \right), \\
\delta^{(2k+1)}_{m_S^2} \sim \delta_{S^{2k+1}} v_S^{2k - 1} \lesssim m_S^2 &\to (c_S)^{ii} \lesssim \left(16 \pi^2 \right)^{\frac{5}{4}} (Y_u^i)^{-\frac{1}{2}} \left( \frac{m_S}{M} \right).
\end{split}
\ee
Because of the loop suppression (and especially in the case of small $Y_u^i$), the limit from the $S^2$ term, which we have also written in Eq.~\eqref{eq:csboundfroms2}, is dominant.

\subsection{Flavor violation}
\label{subsection:flavor}

\begin{figure}[tbp]
\centering
\includegraphics[width=\textwidth]{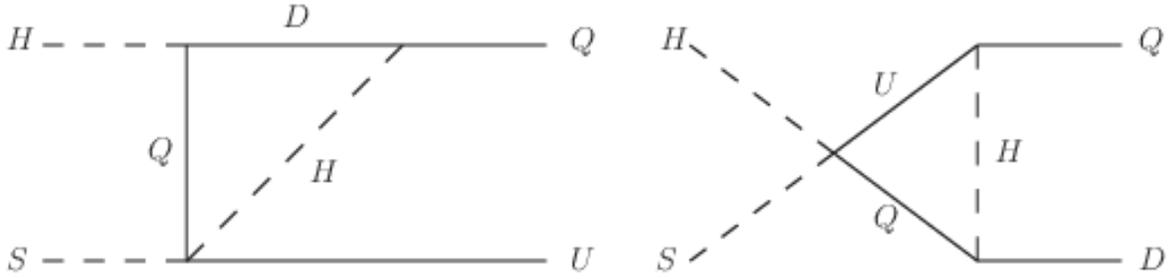}
\caption{
Flavor violation in the up-type (left) and down-type (right) quark sectors, for a coupling that is diagonal in the up-type mass eigenbasis. All flavor violation is provided by the CKM matrix.}
\label{fig:fveft}
\end{figure}
Next, we analyze the flavor violation induced by $c_S$ in Eq.~\eqref{eq:bsmlag}. 
Since the same up-type quark rotations diagonalize $Y_u$ and $c_S$, flavor is preserved by all diagrams involving only the up quarks and the new interaction~\footnote{We consider only couplings of $S$ to a single quark in this work. If $S$ couples to both up-type and down-type quarks, the flavor physics will be similar if the $c_S$ of Eq.\eqref{eq:bsmlag} and its down-type analog are both diagonal in the respective quark mass bases. In this case only, the spurion arguments that follow still hold.}. 
We choose to work in a basis where $Y_u$ is diagonal and $c_S$ has a single diagonal non-zero component. In this basis, the misalignment between the $SU(2)_L$ partners of the up quarks and the left-handed components of the down quark mass eigenstates is given by the CKM matrix, which is in turn defined as $Y_d = \ckm Y_d^D$ where $Y_d^{(D)}$ is the (diagonalized) down Yukawa matrix.
Any flavor violation must come from terms involving the down-type Yukawas. As an example, consider the flavor structure of the operator $S \bar{Q}_L U_R$. By $S$ parity, the coefficient of this operator must be proportional to $c_S$, and its leading component is simply $\frac{c_S v}{\sqrt{2} M}$. We may create a flavor-changing neutral current (FCNC) by writing the simplest contribution to the $S \bar{Q}_L U_R$ term involving $Y_d$. Because $Y_d$ is the only coupling that breaks the down-type quark chiral symmetry, any contribution to the $S \bar{Q}_L U_R$ operator must involve an even number of insertions of $Y_d$. Flavor violation is thus only possible at the expense of two small Yukawas and an off-diagonal CKM element, as in $\left( \ckm Y_d^D (Y_d^D)^\dagger \ckm^\dagger \right) \frac{c_S v}{\sqrt{2} M} S \bar{Q}_L U_R$, in addition to a loop factor as indicated by the diagram in the left panel of Figure~\ref{fig:fveft}.

In addition, even if the new scalar couples only to up-type quarks at tree level, couplings to the down quarks may be induced at loop level. Again from symmetry arguments, the induced $S \bar{Q}_L D_R$ operator must have at least one insertion of each of $c_S$, $Y_u$ and $Y_d$. As above, we need at least one loop; a diagram leading to the operator is shown in the right panel of Figure~\ref{fig:fveft}, and the associated flavor matrix is
$(Y_u^D) c_S^\dagger \ckm Y_d^D$.

Rotating to the down quark mass eigenstate basis, the expected sizes of the off-diagonal elements of the above operator are typically well below the limits from meson mixing for $\mathcal{O}(\mathrm{TeV})$ suppression scales~\cite{Isidori:2010kg,Blankenburg:2012ex}, due to the Yukawa and loop suppressions. Nevertheless, it is instructive to consider the strength of current meson mixing limits. For instance, for $S$ lighter than the kaon mass, effective four-quark operators such as
\be
\left( \frac{1}{16\pi^2 M} \left( \ckm^\dagger (Y_u^D) c_S^\dagger \ckm Y_d^D \right)_{12} \right)^2 \left( \frac{v^2}{2 (m_K^2 - m_S^2)} \right) (\bar{d}_L s_R \bar{d}_L s_R)
\ee
are induced.
If $c_S$ only couples a new scalar to the charm quark, the strongest bound comes from $K$ mixing~\cite{Isidori:2010kg}, and is merely $(c_S)_{22} \lesssim M / (20~\mathrm{GeV})$. On the other hand, if the scalar interaction breaks only the up quark chiral symmetry, the best limit is now even weaker because of the small first generation Yukawas: $D$ mixing gives $(c_S)_{11} \lesssim M /(0.6~\mathrm{GeV})$. We see that our underlying symmetry principle has effectively suppressed flavor-violating interactions, rendering FCNC limits irrelevant. The constraints are much stronger without such a symmetry. For instance, for a scalar coupling to up-type quarks, $(c_S)_{12} / M$ is bounded at the $1/(10^8~\mathrm{GeV})$ level from $D$ mixing bounds.

In addition to models containing a coupling to a particular flavor of quarks, we will also allow for models in which $S$ couples to a single lepton flavor at tree level. To do so in the EFT, we make a straightforward replacement of the quark doublet and singlet with the lepton doublet and singlet, $Q\to L$, $U\to E$, and the interaction of the scalar is
\be
{\cal L}\supset -\frac{c_S}{M}S\bar L_LE_R H+{\rm h.c.}
\label{eq:leptonS}
\ee
In the lepton sector, flavor-specific flavor symmetries can lead to different flavor observables depending on the mechanism responsible for neutrino mass generation. For an interaction of the form $S \bar{L}_L E_R H$, the above treatment can be generalized in the case of Dirac neutrino mass terms, with all flavor violation proportional to small neutrino masses. Alternatively, instead of Dirac neutrino masses, heavy right-handed neutrinos with Majorana masses could be integrated out to produce the effective Weinberg operator $(LH)^2$. In this case, such an operator would give neutrino mixing and be the only source of flavor violation in the lepton sector. It would also induce flavor-violating contributions to $S \bar{L}_L E_R H$, but since the $S$ coupling preserves lepton number, such flavor violation would be suppressed by two powers of the Majorana neutrino mass.

\subsection{Renormalizable models}
\label{subsec:UVcompletions}
The dimension-five operators that we have considered thus far must be resolved at high energies, and in this section we consider fully renormalizable theories that can give rise to the $c_S$ term of Eq.~\eqref{eq:bsmlag}. We may complete the interaction by introducing new vector-like fermions or scalars. In general, both lead to electroweak precision bounds, while the latter are also subject to constraints from mixing with the Higgs. Here we choose to focus on the vector-like fermion completion.

We introduce a vector-like quark doublet with the same gauge charges as $Q_L$ and denote its left- and right-handed components by $Q'_L$ and $Q'_R$, respectively.\footnote{We could equally well have chosen the new vector-like quark to have the same charge as $U_R$, which would not significantly affect the influence of electroweak precision constraints.} Then, the operator with coefficient $c_S$ may be replaced by the Lagrangian
\be
\label{eq:bsmlagcs}
\mathcal{L}_{c_S} = i \bar{Q}'_L \slashed{D} Q'_L + i \bar{Q}'_R \slashed{D} Q'_R - \left( y_S S \bar{Q}_L Q'_R + M \bar{Q}'_R Q'_L + y' \bar{Q}'_L H_c U_R + \mathrm{h.c.} \right).
\ee
The above Lagrangian provides a UV completion of the $S \bar{Q}_L U_R H_c$ operator mediated by the new vector-like quark, and we have deliberately used the same variable $M$ for the vector-like quark mass as for the loop cutoff scale above, assuming that the same physics is responsible for both.

In a similar fashion as above, we may ask about the technical naturalness of the couplings of Eq.~\eqref{eq:bsmlagcs} and the resulting scalar potential. Clearly $y_S$ is natural because it is the only interaction term that breaks $S$ parity. $y'$ is also natural because it breaks a global $Z_2$ symmetry under which the fields $Q', S$ are odd and the remaining fields are even.

From a flavor perspective, Eq.~\eqref{eq:bsmlagcs} motivates the consideration of an enlarged symmetry group $U(4)_Q \times U(3)_U \times U(3)_D \times U(1)_{Q'_R}$, where the left-handed quark flavor group now includes $Q'_L$. Keeping $S$ as a flavor singlet, the couplings $(y_S, M)$ form a $4$ of $U(4)_Q$, while $(Y_u, y')$ fall into the $(4, \bar{3}, 1)$ bifundamental representation. Our flavor-specific flavor principle may be restated in terms of the symmetry breaking pattern of the new couplings. For instance, the up-specific structure of Eq.~\eqref{eq:cS-break} may be written as the hypothesis that the new couplings break the full symmetry group to $U(1)_{u+q'} \times U(2)_{ctL} \times U(2)_{ctR} \times U(3)_D$, where the former symmetry corresponds to a simultaneous chiral rotation of the up quark and new vector-like quark.

However, given the presumably different natures of the couplings in each $4$ of the new $U(4)_Q$ above (as hinted by, e.g., their varying $S$ shift symmetry properties), we choose to analyze flavor through the standard SM flavor group. Under the usual $U(3)^3$ of the SM quark sector, the vector-like quark is simply a flavor singlet, and the couplings of Eq.~\eqref{eq:bsmlagcs} have the flavor structure
\begin{align}
y_S &\sim (3, 1, 1), \nonumber \\
M &\sim (1, 1, 1), \\
y' &\sim (1, \bar{3}, 1). \nonumber
\end{align}
The up-specific principle is now the statement that the new couplings break the $U(3)_Q \times U(3)_U \times U(3)_D \times U(1)_{Q'_L} \times U(1)_{Q'_R}$ symmetry down to the same $U(1)_{u+q'} \times U(2)_{ctL} \times U(2)_{ctR} \times U(3)_D$ as before. Given this assumption, if we work in the basis where $Y_u$ is diagonal, $y_S$ and $y'$ can each have only one non-zero element, and as in the effective theory all flavor violation comes from $Y_d$. Now let us consider the sizes of the flavor-violating interactions $S \bar{Q}_L U_R$ and $S \bar{Q}_L D_R$, as we did in Sec.~\ref{subsection:flavor} for the effective theory. The simplest way to obtain non-trivial flavor structure in a term breaking the $S$ shift symmetry is to use the combination $y_S y'$ with the down-type Yukawas. While other terms are possible, they involve higher powers of the new couplings, so to leading order the FCNC limits are the same as in Sec.~\ref{subsection:flavor} with $c_S \to y_S y'$.

Focusing on the scalar potential, we note that for even $n$, there is now a \emph{one-loop} correction to $S^n$ involving $n$ insertions of $y_S$ to make a loop of $Q$ and $Q'$. For odd $n$, there is no one-loop contribution, but we may add a loop involving a Higgs and containing the vector-like mass $M$ as well as the couplings $y'$ and $Y_u$. We then have
\be
\begin{split}
\label{eq:sncorrcs}
\delta_{S^{2k}} &\sim \frac{\Tr (y_S^\dagger y_S)^k}{16\pi^2} M^{4 - 2k}, \\
\delta_{S^{2k + 1}} &\sim \frac{\Tr (y_S^\dagger y_S)^k y_S^\dagger y' Y_u^\dagger}{(16\pi^2)^{2}} M^{4 - (2k + 1)}.
\end{split}
\ee
For sufficiently high cutoff scales, we may again ignore mixed scalar potential terms involving both $S$ and $H$. Note that unlike the non-renormalizable model we considered before, there is a one-loop $S$ mass correction. It goes as
\be
\label{eq:s2boundcs}
\delta m_S^2 \sim \frac{\Tr y_S^\dagger y_S}{16\pi^2} M^2
\ee
so the bound on the elements of $y_S$ is
\be
\label{eq:ysboundfroms2}
(y_S)^{ij} \lesssim (4\pi) \frac{m_S}{M} \approx (3 \times 10^{-4}) \left( \frac{m_S}{0.1~\mathrm{GeV}} \right) \left( \frac{5~\mathrm{TeV}}{M} \right).
\ee
While $y'$ does not appear in the above expression, it does give a one-loop correction to the Higgs mass. We require the Higgs mass correction to be no larger than $v$ itself, yielding the relatively weaker bound
\be
\label{eq:ypboundfromh2}
(y')^{ij} \lesssim (4\pi) \frac{v}{M} \approx (6 \times 10^{-1}) \left( \frac{5~\mathrm{TeV}}{M} \right).
\ee
The product of the limits in Eqs.~\eqref{eq:ysboundfroms2} and~\eqref{eq:ypboundfromh2} may be compared with that for the non-renormalizable theory in Eq.~\eqref{eq:csboundfroms2}. We see that in the full theory, the constraint on the size of the effective dimension-five operator is stronger by a factor $v / M$.

The corrections to $S^n$ are suppressed by fewer loops in the full theory than in the non-renormalizable one. However, recall that the limit on $y_S$ from naturalness of the $S$ mass is more stringent than the limit on $c_S$, by a factor of $4 \pi$ or ``half'' a loop factor. Consequently, the behavior of the $S$ potential in the presence of the corrections of Eq.~\eqref{eq:sncorrcs} is similar to that in the non-renormalizable theory with the corrections of Eq.~\eqref{eq:sncorr}. The $S$ and $S^2$ terms largely determine the potential and set the $S$ vev, which is as in Eq.~\eqref{eq:svev} with the replacement $c_S \to y_S^\dagger y'$. Because the constraint on the product $y_S^\dagger y'$ from naturalness is mildly stronger than that on $c_S$ by a factor $v/M$, the natural size of the $S$ vev tends to be slightly smaller in the fully renormalizable theory.

\subsection{$CP$ violation}
\label{subsection:cpv}

Finally, we discuss the behavior of the scalar interaction with fermions under charge conjugation, $C$, and a parity transformation, $P$. For definiteness, we assume that the scalar couples only to one flavor of fermion, in particular the $u$ quark here. After electroweak symmetry breaking the relevant interaction is
\be
{\cal L}_{\rm int}=-\frac{c_S v}{\sqrt2 M}S\bar u_Lu_R-\frac{c_S^\ast v}{\sqrt2 M}S\bar u_Ru_L=-\frac{v}{\sqrt2 M}S\bar u\left[{\rm Re}(c_S)+i{\rm Im}(c_S)\gamma^5\right]u.
\label{eq:LintSuu}
\ee
Once the $u$ quark mass is made real by a chiral rotation there is no longer enough freedom to rephase the fields in ${\cal L}_{\rm int}$ because $S$ is a real scalar field. Therefore the phase of the coupling $c_S$ is physical. Under $P$, $\bar u_Lu_R\leftrightarrow \bar u_Ru_L$. Thus, $P$ can be made a good symmetry if $c_S$ is purely real or purely imaginary by taking $S\to S$ or $S\to-S$, respectively, under $P$. Note that since $C$ is conserved by ${\cal L}_{\rm int}$, $P$ conservation implies $CP$ conservation. In the case of purely real or purely imaginary $c_S$, all $CP$ violation comes from the CKM matrix, leading to a large suppression by light quark Yukawas. Moreover, if $S$ is a pseudoscalar, all $S^n$ potential terms with $n$ odd are forbidden by $P$ invariance, and in particular $S$ does not acquire a vev.

For a generic value of the phase of $c_S$, however, $CP$ is not a good symmetry of ${\cal L}_{\rm int}$, leading to $CP$ violating processes involving $S$. In our example of a coupling to $u$ quarks, a neutron electric dipole moment (EDM) develops and therefore the strong experimental upper limit on the neutron EDM can be used to constrain the size of the coupling. Below, we estimate the neutron EDM that results when the $S$-$u$-$u$ coupling has a nontrivial phase.

\begin{figure}[tbp]
\centering
\includegraphics[width=0.5\textwidth]{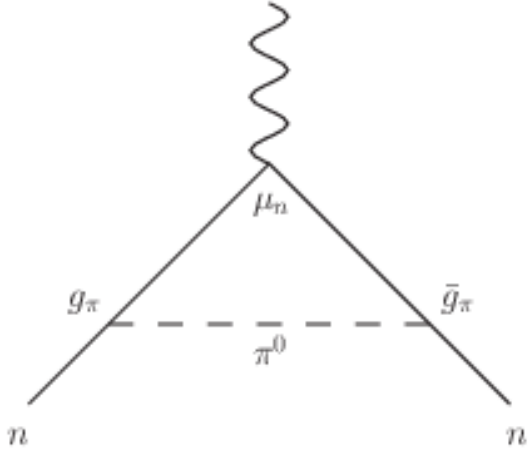}
\caption{One-loop contribution to the neutron EDM in the presence of a $CP$-violating $\pi^0$-neutron coupling $\bar g_\pi$.}
\label{fig:nEDM}
\end{figure}

An imaginary $c_S$ in Eq.~(\ref{eq:LintSuu}) causes the $S$ to mix with pseudoscalar mesons. The mixing angle with the $\pi^0$, for instance can be estimated to be
\be
\begin{aligned}
\theta_{\pi S}&\simeq\frac{f_{\pi}}{\sqrt 2\left(m_u+m_d\right)}\,\frac{m_\pi^2}{m_S^2-m_\pi^2}\,\frac{{\rm Im}(c_S) v}{\sqrt2 M}
\\
&\simeq6\times10^{-3}{\rm Im}(c_S)\left(\frac{1~\rm GeV}{m_S}\right)^2\left(\frac{5~\rm TeV}{M}\right).
\end{aligned}
\label{eq:thetaSpi}
\ee
In the last step we have assumed that $m_S\gg m_\pi$. The real part of $c_S$ leads to $S$ developing a scalar coupling to nucleons in the low energy effective theory. In particular, its coupling to neutrons is
\be
{\cal L}_{\rm eff}\supset-\frac{{\rm Re}(c_S) v}{\sqrt2 M}\,\frac{m_nf_u^n}{m_u}\,S\bar n n,
\label{eq:Snn}
\ee
where we have used the matrix element $\langle n|\bar uu|n\rangle=(m_nf_u^n)/m_u$ with $f_u^n\simeq0.011$~\cite{Belanger:2013oya}. Now, in addition to its usual $CP$-conserving coupling to nucleons, $g_\pi\simeq13.4$, in the presence of (\ref{eq:thetaSpi}) and (\ref{eq:Snn}) the $\pi^0$ obtains a $CP$-violating coupling to neutrons, $\bar g_\pi$,
\be
{\cal L}_{\rm eff}\supset- \frac{1}{2}\pi^0\bar n\left(\bar g_\pi+ig_\pi\gamma^5\right)n,
\ee
with
\be
\begin{aligned}
\bar g_\pi&\simeq\sqrt2\,\frac{{\rm Re}(c_S) v}{M}\,\frac{m_nf_u^n}{m_u}\,\theta_{\pi S}
\\
&\simeq2.5\times10^{-3}\left|c_S\right|^2\left(\frac{\sin 2\beta}{2}\right)\left(\frac{1~\rm GeV}{m_S}\right)^2\left(\frac{5~\rm TeV}{M}\right)^2,
\end{aligned}
\ee
and $\beta\equiv\arg c_S$. The $CP$-violating coupling leads to a neutron EDM. A simple estimate of this can be obtained by evaluating a one-loop diagram, shown in Fig.~\ref{fig:nEDM}, with a pion loop and a photon coupled to the neutron through its magnetic dipole moment, $\mu_n$.
Cutting this loop off at the neutron mass gives a simple expression for the EDM,
\be
\left|d_n\right|\sim\frac{\bar g_\pi g_\pi}{32\pi^2}\left|\mu_n\right|\simeq1\times10^{-18}e\,{\rm cm}\left|c_S\right|^2\left|\frac{\sin 2\beta}{2}\right|\left(\frac{1~\rm GeV}{m_S}\right)^2\left(\frac{5~\rm TeV}{M}\right)^2.
\ee
Requiring that this is less than the experimental upper limit of $0.3\times 10^{-25}e\,{\rm cm}$~\cite{Afach:2015sja} results in a limit of
\be
\left|c_S\right|\times\left|\frac{\sin 2\beta}{2}\right|^{1/2}\lesssim2\times10^{-4} \left(\frac{m_S}{1~\rm GeV}\right)\left(\frac{M}{5~\rm TeV}\right),
\ee
or, in terms of the $S$ coupling to $u$ quarks, $g_S^{uu}=c_Sv/\sqrt2M$,
\be
\left|g_S^{uu}\right|\times\left|\frac{\sin 2\beta}{2}\right|^{1/2}\lesssim6\times10^{-6} \left(\frac{m_S}{1~\rm GeV}\right).
\ee

In addition to the $\pi^0$-$S$ mixing effect outlined above, there is another contribution to the neutron EDM when $S$ develops a vev. If there is an imaginary $c_S$ in such a case, then this leads to a phase for the $u$ quark mass. This contributes to the physical $\theta$ angle of QCD through $\bar\theta=\theta-\arg\det M_q$ where $M_q$ is the light quark mass matrix. In the presence of an $S$ vev, $\langle S\rangle=v_S$, this is
\be
\bar\theta=-\tan^{-1}\frac{{\rm Im}(c_S)v}{\sqrt2 m_u}\,\frac{v_S}{M}.
\ee
A nonzero $\bar\theta$ contributes to the neutron EDM~\cite{Crewther:1979pi} and the present limit can be interpreted as an upper limit on the magnitude of $\bar\theta$ of about $10^{-10}$ or
\be
{\rm Im}(c_S)\lesssim 10^{-10}\,\frac{\sqrt2 m_u}{v}\,\frac{M}{v_S}.
\label{eq:edmlimitfromvev}
\ee
The real part of $c_S$ contributes to an $S$ tadpole which induces an $S$ vev as described in Sec.~\ref{sec:potential}. Using the expected vev from Eq.~(\ref{eq:svev}) in this expression, we have
\be
\left|c_S\right|\times\left|\frac{\sin 2\beta}{2}\right|^{1/2}\lesssim 3\times10^{-7}\left(\frac{m_S}{1~\rm GeV}\right)\left(\frac{5~\rm TeV}{M}\right),
\ee
which implies for the $S$-$u$-$u$ coupling,
\be
\left|g_S^{uu}\right|\times\left|\frac{\sin 2\beta}{2}\right|^{1/2}\lesssim1\times10^{-8}\left(\frac{m_S}{1~\rm GeV}\right)\left(\frac{5~\rm TeV}{M}\right)^2.
\ee
For the set of parameters we have normalized on, the limit from this contribution is a couple of orders of magnitude stronger than the limit from $\pi^0$-$S$ mixing. However, this limit depends on there being an $S$ vev which one could imagine is tuned away while the mixing contribution remains robust.

In any case, for ${\cal O}(1)$ phases of $c_S$, the neutron EDM provides a strong constraint on the size of the coupling to light quarks. Therefore, to obtain an appreciable coupling, we are led to consider UV theories in which $CP$ is a good symmetry of the $S$-$u$-$u$ coupling, taking $S$ to be either scalar or pseudoscalar.

We can also ask about $CP$ violation in the case of a coupling of the scalar to leptons through the interaction of Eq.~\eqref{eq:leptonS}. First, consider a coupling just to electrons, i.e. we can write the coupling matrix in the mass basis as $c_S\delta^{i1}\delta^{j1}$. One can then write down a one-loop contribution to the electron EDM,
\be
\begin{aligned}
\left|d_e\right|&\sim \frac{\left|g_S^{ee}\right|^2}{8\pi^2}\left|\frac{1}{2}\sin 2\beta\right|\left|\mu_e\right|\frac{m_e^2}{m_S^2}\log\frac{m_S^2}{m_e^2}
\\
&\simeq 6.4\times10^{-19}e\,{\rm cm}\left|g_S^{ee}\right|^2\left|\frac{1}{2}\sin 2\beta\right|\left(\frac{1~\rm GeV}{m_S}\right)^2\left(\frac{\log m_S^2/m_e^2}{10}\right).
\end{aligned}
\ee
where $\mu_e$ is the electron magnetic dipole moment, $g_S^{ee}=c_Sv/\sqrt2 M$, and $\beta$ is the phase of $c_S$. This must be less than the experimental upper limit on the electron EDM of $0.87\times10^{-28}e\,{\rm cm}$~\cite{Baron:2013eja}, which means that
\be
\left|g_S^{ee}\right|\left|\frac{1}{2}\sin 2\beta\right|^{1/2}\lesssim 1.2\times10^{-5}\left(\frac{m_S}{1~\rm GeV}\right)\left(\frac{10}{\log m_S^2/m_e^2}\right)^{1/2}.
\ee
In the case of a leading coupling of $S$ to other lepton flavors (or quarks) that is $CP$-violating, the constraint from the electron EDM is much weaker, since the induced electron EDM occurs only at three loops.

Besides providing insights from technical naturalness, $CP$-like symmetries can be instrumental in constructing phenomenologically viable theories. In the next section, we will show an application involving dark matter, where taking a pseudoscalar $S$ naturally avoids direct detection bounds.

\section{Applications}
\label{sec:applications}

We now turn to applications of the framework described in the previous section. In particular, we consider a model of a scalar which mediates interactions with DM and preferentially couples to up quarks. This is distinct from typical scalar simplified DM models, which usually have very small couplings to first-generation fermions (for some recent work, see, e.g.,~\cite{Baek:2011aa,Buckley:2014fba,Buchmueller:2015eea,Abdallah:2015ter,Abercrombie:2015wmb,Boveia:2016mrp,Englert:2016joy,Bauer:2016gys,Baek:2017vzd}. We also consider the theory of a light scalar which couples only to muons. Such a state could explain the currently measured value of the muon $g - 2$ without running afoul of constraints from electron couplings.

\subsection{Up-specific scalar mediated dark matter}

First, we consider a scalar that couples to the up quark, corresponding to $c_S^{ij} = c_u \delta^{i1} \delta^{j1}$ (see also~\cite{Alanne:2017oqj}). For a GeV-scale scalar with a cutoff at several TeV, the natural value of the physical renormalizable $S \bar{u} u$ coupling is relatively small. Such couplings are typically below the $\mathcal{O}(1)$ limits on light dijet resonances from UA2~\cite{Alitti:1990kw}, which LHC searches are only now starting to improve~\cite{Sirunyan:2017nvi}. For scalar masses above 100 GeV, collider dijet bounds put more severe limits on natural couplings~\cite{Abe:1997hm,Aaltonen:2008dn,Dobrescu:2013coa,Aad:2014aqa,Khachatryan:2016ecr,Sirunyan:2017nvi}. For scalars below roughly 1 GeV, intensity frontier experiments must be taken into account, as well as astrophysical bounds, which requires the evaluation of non-perturbative hadronic and nuclear effects. Thus we here choose to focus on the intermediate region.

We introduce a new fermionic Dirac DM particle $\chi$ with vector-like mass $m_\chi$ and assume that it has a coupling to $S$. That is, we consider the interactions
\be
\label{eq:dmlag}
\mathcal{L}_{\mathrm{hidden}} = i \bar{\chi}_L \slashed{\partial} \chi_L + i \bar{\chi}_R \slashed{\partial} \chi_R - \left(m_\chi \bar{\chi}_L \chi_R + y^\chi_S S \bar{\chi}_L \chi_R + \mathrm{h.c.}\right),
\ee
and assume that $\chi$ annihilation to up quarks is responsible for setting the relic abundance. The phases of the $S$ couplings to the SM and DM, $g^{uu}_S \equiv \frac{c_S v}{\sqrt{2} M}$ and $y^\chi_S$, affect the signatures of the theory as real and imaginary couplings lead to different phenomenology. We now proceed to describe the potential signatures in terms of the possible coupling choices.

We begin by recalling from Sec.~\ref{subsection:cpv} that if $g^{uu}_S$ contains both real and imaginary components, a neutron EDM arises which is strongly constrained by experiment. Consequently, we consider only purely real or imaginary $g^{uu}_S$. Now, we examine the $CP$ implications of the coupling between $S$ and DM. If $g^{uu}_S$ and $y^\chi_S$ have the same phase, then there is still a good $CP$ symmetry, and no EDM is generated. On the other hand, if $y^\chi_S$ has a non-trivial phase relative to $g^{uu}_S$, there is no unique assignment of the $S$ parity that allows the full action to be preserved under $CP$.

When $g^{uu}_S$ is imaginary, a real component of $y^\chi_S$ is dangerous in this regard because it leads to a one-loop scalar vev
\be
v_S \sim \frac{\mathrm{Re}\,y^\chi_S\,m_\chi}{16 \pi^2} \left( \frac{M}{m_S} \right)^2
\ee
where we have used the same cutoff scale $M$ as in Sec.~\ref{sec:symmetries}. Together, the vev and $g^{uu}_S$ induce a contribution to the QCD $\theta$ angle which is bounded as in Eq.~\eqref{eq:edmlimitfromvev}, giving
\be
\mathrm{Im}\,g^{uu}_S\,\mathrm{Re}\,y^\chi_S \lesssim (10^{-10}) (16 \pi^2) \frac{m_u}{m_\chi} \left( \frac{m_S}{M} \right)^2
\ee
The limit above essentially enforces that if the $S$ coupling to up quarks is imaginary in this model, then $S$ should transform as a pseudoscalar in its hidden sector interactions as well.

The case of real $g^{uu}_S$ is less constrained by EDM searches. To see this, we first note that if $S$ is assigned even parity, then all $CP$ violation comes from the imaginary component of $y^\chi_S$. Consequently, any $CP$-violating operator must be proportional to an odd number of powers of $y^\chi_S$. To obtain $CP$ violation involving the SM fields only, we thus need a $\chi$ loop. Since the loop involves $\gamma^5$, it follows that there must be at least five scalars attached to it. No EDM in the SM can arise, then, below five loops.
In summary, for real $g^{uu}_S$, $y^\chi_S$ is not barred from having an arbitrary phase by EDM searches alone.

\begin{figure}[tbp]
\centering
\includegraphics[width=\textwidth]{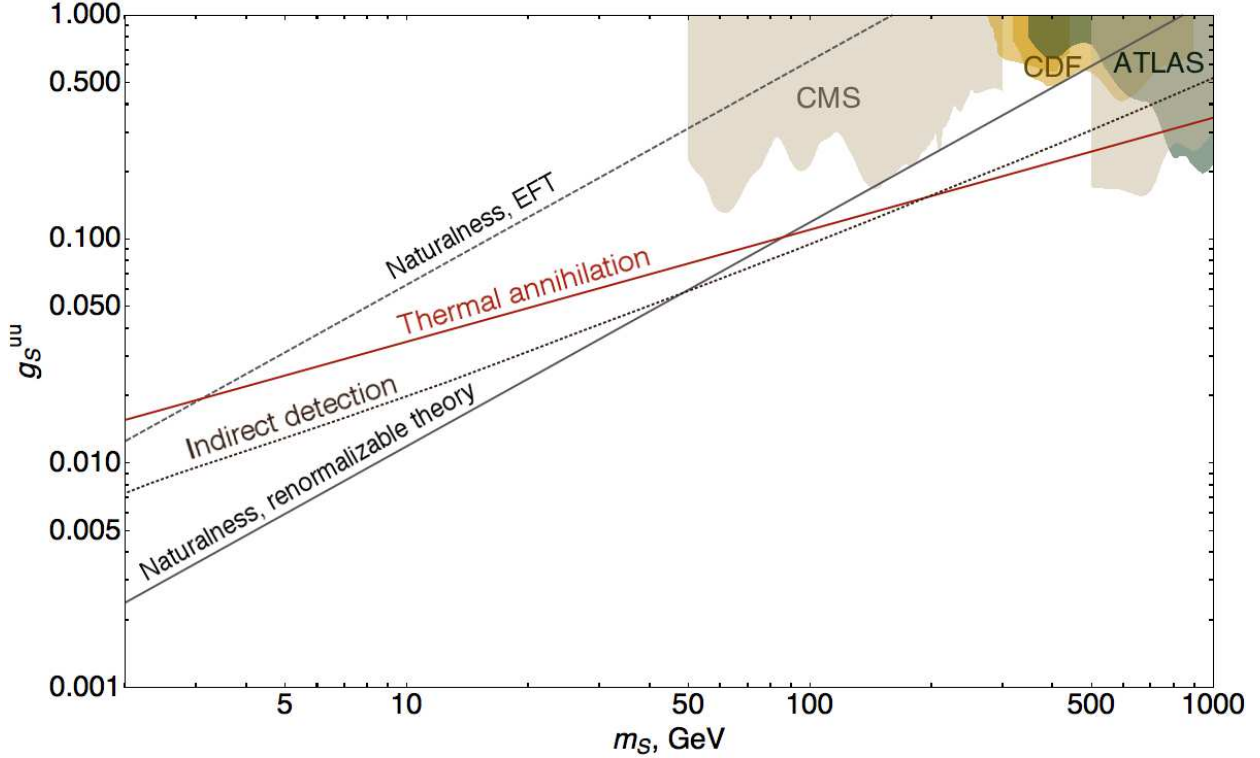}
\caption{\label{fig:sqq} 
Constraints on a light pseudoscalar coupling to quarks and DM in the $m_S - g_S^{uu}$ plane. 
The red line indicates where the annihilation cross section is equal to the canonical thermal relic value, $\langle \sigma v \rangle = 3 \times 10^{-26}$ cm$^3$ s$^{-1}$. 
The region above the brown dotted line labeled is excluded by Fermi-LAT gamma-ray observations from dwarf spheroidal satellite galaxies~\cite{Ackermann:2015zua}. 
Also displayed are bounds from dijet searches at the Tevatron~\cite{Abe:1997hm,Aaltonen:2008dn,Dobrescu:2013coa} and LHC~\cite{Aad:2014aqa,Khachatryan:2016ecr,Sirunyan:2017nvi}, rescaled using MadGraph 5~\cite{Alwall:2014hca}. 
Finally, the region below the black dashed line (black solid line) is natural according to the EFT (renormalizable model) criterion presented in Eq.~(\ref{eq:csboundfroms2}) (Eqs.~(\ref{eq:ysboundfroms2},\ref{eq:ypboundfromh2})).
A 2 TeV cutoff scale is assumed.
}
\end{figure}
 
Now, independently of $CP$ violation, an imaginary component of $y^\chi_S$ can be problematic for indirect detection. This can be seen from the DM annihilation cross section, which to second order in the DM relative velocity $v$ is~\cite{Berlin:2014tja}
\begin{multline}
\sigma v (\bar{\chi} \chi \to \bar{u} u) \approx \frac{3\,(\mathrm{Im}\,y^\chi_S)^2 |g^{uu}_S|^2 m_\chi^2}{2 \pi \left( m_S^2 - 4 m_\chi^2 \right)^2}\ +\\
v^2 \left( \frac{3 |g^{uu}_S|^2 m_\chi^2}{8\pi \left( m_S^2 - 4 m_\chi^2 \right)^3} \right) \left( (\mathrm{Im}\,y^\chi_S)^2 (m_S^2 + 4 m_\chi^2) + (\mathrm{Re}\,y^\chi_S)^2 \left( m_S^2 - 4 m_\chi^2 \right) \right).
\label{eq:sigmav}
\end{multline}
In the above we have neglected the final state quark mass and ignored hadronization effects, which should be a good approximation for $m_\chi \gtrsim \Lambda_{\mathrm{QCD}}$. We see that annihilation is $s$-wave ($p$-wave) for imaginary (real) $y^\chi_S$, regardless of the phase of the scalar-SM coupling. For $m_\chi \lesssim 100~\mathrm{GeV}$, $s$-wave DM annihilation to up quarks which produces the observed relic density with a standard thermal cosmology is in tension with Fermi-LAT observations of dwarf spheroidal galaxies~\cite{Ackermann:2015zua}.
However, strong limits in the case of imaginary $y^\chi_S$ from DM annihilation can be evaded if the DM abundance is set by an early $\chi$-$\bar{\chi}$ asymmetry.
In Figure~\ref{fig:sqq}, we thus show the parameter space of a scalar coupling to up quarks and DM with imaginary couplings. We plot the naturalness bounds of the previous section in both the effective theory with the operator $S \bar{Q}_L H U_R$ and a possible ultraviolet completion with vector-like quarks having the same SM gauge charges as the left-handed quark doublet. For comparison, we choose a fixed mass ratio $m_{\chi} / m_S = 3/4$ and show the area where the annihilation cross section is the standard thermal value $\langle \sigma v \rangle = 3 \times 10^{-26}~\mathrm{cm}^3/\mathrm{s}$, assuming that $|y^\chi_S| = \frac{c_S v}{\sqrt{2} M}$, i.e. that the $S \bar{u} u$ and $S \bar{\chi} \chi$ couplings are equal. The region above the dotted indirect detection line requires additional physics such as the aforementioned DM asymmetry to be viable. A small window remains for the thermal DM scenario at masses of a few hundred GeV, above which dijet limits become constraining.

The only remaining case is that of real $g^{uu}_S$ and $y^\chi_S$. However, real $g^{uu}_S$ \emph{and} $y^\chi_S$ would lead to an unsuppressed spin-independent direct detection cross section
\be
\sigma^N_{\mathrm{SI}} = \frac{m_\chi^2 m_N^4}{\pi m_u^2 m_S^4 (m_\chi + m_N)^2} (f^N_u)^2 (g^{uu}_S y^\chi_S)^2,\ N = n,p
\ee
where $f^p_u \approx 0.015$ and $f^n_u \approx 0.011$ are the same form factors we used in Sec.~\ref{subsection:cpv}~\cite{Belanger:2013oya}. 
The resulting cross section is tightly limited~\cite{Aprile:2017iyp,Petricca:2017zdp}. For $g^{uu}_S = y^\chi_S$, the coupling that would be necessary to obtain the observed relic DM abundance is already excluded.

We thus see that an up-specific scalar is rather constrained as a DM mediator, by a combination of direct detection, indirect detection, and EDM searches. With standard assumptions about cosmology, the only viable scenarios are a pseudoscalar that mediates interactions between DM and up quarks for relatively heavy $\chi$, or the alternative case of secluded annihilation where DM annihilates to $S$ itself.

\subsection{Muon-specific EFT}

Here we show another application of our formalism to a scalar that couples solely to muons at tree level. A muon-specific scalar could account for the discrepancy between the experimentally measured value of the anomalous magnetic dipole moment of the muon and its theoretical prediction~\cite{Gninenko:2001hx,Fayet:2007ua,Pospelov:2008zw,Davoudiasl:2012ig}, which is currently at the level of 3-4 standard deviations~\cite{Roberts:2010cj,Hagiwara:2011af}. The usual MFV choice is to postulate a new scalar with leptonic coupling strengths proportional to $m_\ell$, which is constrained from the electron couplings~\cite{Batell:2016ove}. By contrast, a strictly muon-specific scalar can easily be long-lived for $m_S < 2 m_\mu$, leading to late decays with potential signatures at fixed-target experiments~\cite{Chen:2017awl}. In this regime, the induced loop-level photon coupling can still lead to appreciable limits from beam dumps and supernovae.

We begin with an analysis of the EFT that leads to a scalar coupled to muons. As we mentioned in Secs.~\ref{subsection:flavor} and \ref{subsection:cpv}, modifying the interactions to involve leptons involves a straightforward replacement of the quark doublet and singlet with the lepton doublet and singlet. The relevant interactions involving the Higgs and the new scalar are then
\be
-{\cal L}_{\rm int}=\bar L_LY_\ell E_R H+\frac{c_S}{M}S\bar L_L E_R H+{\rm h.c.}
\label{eq:leptonEFT}
\ee
As in the case of an up-specific coupling, we assume that $c_S$ and $Y_\ell$ are aligned, and that in the basis where $Y_\ell$ is diagonal, $Y_\ell\propto{\rm diag}(m_e,m_\mu,m_\tau)$, $c_S$ takes the form $c_S={\rm diag}(0,(c_S)^{22},0)$. As in the case of quarks analyzed above, $Y_\ell$ breaks the global lepton family symmetry
$U(3)_L\times U(3)_E\to U(1)_e\times U(1)_\mu\times U(1)_\tau$
while $c_S$ breaks $U(3)_L\times U(3)_E\to U(1)_\mu\times U(2)_{e\tau L}\times U(2)_{e\tau R}$. Crucially, to avoid FCNCs, the $U(1)_\mu$ subgroups left unbroken by $Y_\ell$ and $c_S$ must coincide. 

\begin{figure}[tbp]
\centering
\includegraphics[width=\textwidth]{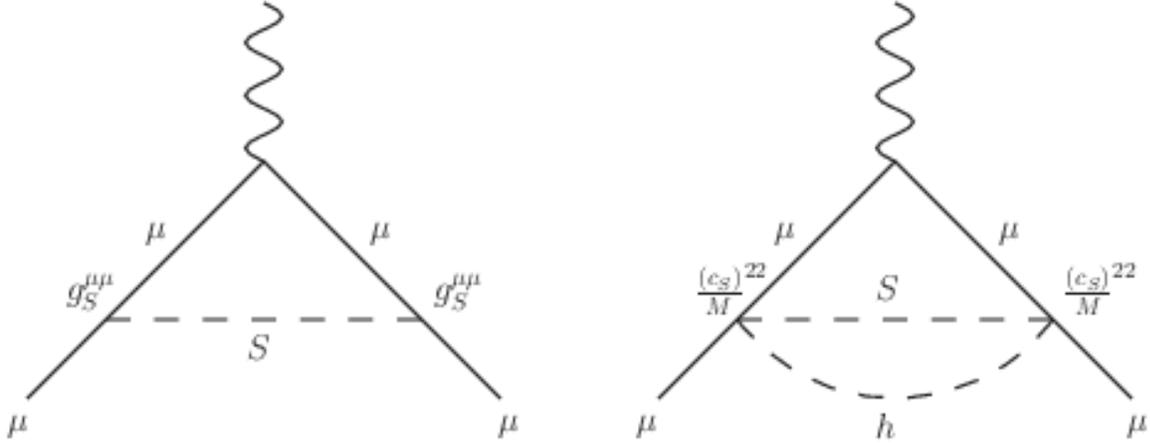}
\caption{One- and two-loop contributions to the muon anomalous magnetic moment in the effective theory of Eq.~\eqref{eq:leptonEFT}. The two-loop contribution is related to that of the one-loop by roughly the factor $M^2/(8\pi^2v^2)$, cf. Eq.~\eqref{eq:2loopratio}.}
\label{fig:g-2}
\end{figure}

After electroweak symmetry breaking, the interactions of Eq.~(\ref{eq:leptonEFT}) lead to a coupling of the scalar to muons,
$-{\cal L}_{\rm int}\supset S\bar\mu \left( \mathrm{Re}\,g_S^{\mu\mu} + i \mathrm{Im}\,g_S^{\mu\mu} \gamma^5 \right) \mu$ with
\be
g_S^{\mu\mu}=\frac{(c_S)^{22} v}{\sqrt{2} M}.
\label{eq:gmumuEFT}
\ee
$S$ exchange, as seen on the left of Fig.~\ref{fig:g-2}, contributes to the muon's magnetic moment with a value proportional to the square of this coupling~\cite{Leveille:1977rc},
\be
\Delta a_\mu=\frac{1}{8\pi^2}\int_0^1dx\frac{\left(1-x\right)^2\left(\left(1+x\right) \left(\mathrm{Re}\,g_S^{\mu\mu}\right)^2 - \left(1-x\right) \left(\mathrm{Im}\,g_S^{\mu\mu}\right)^2\right)}{\left(1-x\right)^2+x\left(m_S/m_\mu\right)^2}.
\label{eq:g-2_1loop}
\ee
A pseudoscalar coupling to muons gives a negative contribution to $\Delta a_\mu$, worsening the discrepancy. This is partially why we do not consider the derivatively coupled operator proportional to $d_S$ in Eq.~(\ref{eq:bsmlag}), and we will henceforth assume that $g_S^{\mu\mu}$ is real.
As originally pointed out in Ref.~\cite{Kinoshita:1990aj}, a scalar with mass $m_S\lesssim m_\mu$ that couples to muons with SM Higgs strength, $g_S^{\mu\mu}=m_\mu/v\sim4\times10^{-4}$, gives a contribution to $\Delta a_\mu$ that is of roughly the right size to explain the discrepancy, $\Delta a_\mu\sim3\left(g_S^{\mu\mu}/4\pi\right)^2=3\left(m_\mu/4\pi v\right)^2=35\times 10^{-10}$.

In addition to the one-loop $S$ exchange contribution to $\Delta a_\mu$, there is a two-loop contribution from the exchange of $S$ and a Higgs shown on the right of Fig.~\ref{fig:g-2}. The ratio of this contribution to the one-loop value is roughly
\be
\frac{(\Delta a_\mu)_{\rm 2-loop}}{(\Delta a_\mu)_{\rm 1-loop}}\sim \frac{M^2}{8\pi^2 v^2},
\label{eq:2loopratio}
\ee
where we have again cut the loop momenta off at $M$. In other words, for $M\lesssim 4\pi v/\sqrt{2}=2~\rm TeV$, the two-loop contribution to $\Delta a_\mu$ can be neglected in comparison to the one-loop value.

In Fig.~\ref{fig:sll}, we show in light red the range of couplings $g_S^{\mu\mu}$ as a function of $m_S$ that bring the measurement and expectation for $(g-2)_\mu$ to within $2\sigma$, using the one-loop expression of Eq.~(\ref{eq:g-2_1loop}) for $\Delta a_\mu$. Above, we also show in dark red the region where the new scalar's contribution to $(g-2)_\mu$ would bring the muon magnetic moment to $5\sigma$ above its measured value.

As described in Sec.~\ref{subsec:EFTnaturalness}, there are corrections to the $S^2$ operator at two loops and to the $S^2H^2$ operator at one loop. Requiring that the shifts $\delta m_S^2$ from each of these operators (after $H$ attains its vev $v/\sqrt2$) are not larger than $m_S^2$ itself leads to an upper bound on the coupling $(c_S)^{22}$. We can then turn this into an upper bound on the coupling of $S$ to muons,
\be
\begin{aligned}
g_S^{\mu\mu}&\lesssim{\rm min}\left[\frac{16\pi^2}{\sqrt2}\frac{m_S v}{M^2},4\pi\frac{m_S}{M}\right]
\\
&\simeq{\rm min}\left[1\times10^{-2}\left(\frac{m_S}{0.1~\rm GeV}\right)\left(\frac{500~\rm GeV}{M}\right)^2,\ 3\times10^{-3}\left(\frac{m_S}{0.1~\rm GeV}\right)\left(\frac{500~\rm GeV}{M}\right)\right].
\end{aligned}
\label{eq:gSlimEFT}
\ee
In Fig.~\ref{fig:sll}, we show the naturalness limit on the coupling
as a solid black line, where for our cutoff choice $M=500~\rm GeV$ the limit comes from the $S^2 H^2$ operator.

\subsubsection{UV completion}
In Eq.~(\ref{eq:leptonEFT}), the scalar interacts with leptons through a dimension-five operator. As we saw in Sec.~\ref{subsec:UVcompletions}, a UV complete theory may introduce additional restrictions on the couplings and masses if we wish to have a natural theory. As before, we take a simple UV completion with a vector-like weak $SU(2)$ doublet $L'$ that has the same quantum numbers at the SM lepton doublet $L_L$. The relevant interactions are
\be
\label{eq:UVmuonlag}
-\mathcal{L}\supset M \bar{L}'_L L'_R+y_S S \bar{L}_L L'_R + y' \bar{L}'_L H E_R+{\rm h.c.}
\ee
Our assumption of a muon-specific coupling means that the flavor structures of $y_S$ and $y^\prime$ are such that only the second generation SM lepton fields $\mu_L$ and $\mu_R$ are involved in the interaction with the vector-like lepton and Higgs. In what follows we therefore drop the flavor indices on $y_S$ and $y^\prime$.

In this theory, $S$ and the Higgs receive one-loop corrections to their (squared) masses. If we require that these are no larger than the squared masses themselves, we get upper bounds on the couplings
that are analogous to Eq.~\eqref{eq:ysboundfroms2} and~\eqref{eq:ypboundfromh2}.

Using hats to denote mass eigenstates, after electroweak symmetry breaking, $L_L^\prime=\hat L_L^\prime$ pairs up with $\hat L_R^\prime=\cos\theta L_R^\prime+\sin\theta \mu_R$ to form a Dirac fermion with mass $\sqrt{M^2+{y^\prime}^2v^2/2}$ where the mixing angle is given by $\tan\theta=y^\prime v/\sqrt2 M$.
The orthogonal combination, $\hat \mu_R^\prime=\cos\theta \mu_R-\sin\theta L_R^\prime$, marries $\mu_L =\hat \mu_L$ to form the light fermion that we identify with the muon. The couplings of the scalar can then be expressed in terms of mass eigenstates,
\be
y_S S \bar{\mu}_L L'_R+{\rm h.c.}=-y_S\sin\theta S \bar{\hat\mu}\hat \mu+\frac{y_S}{2}\cos\theta S \left[\bar{\hat\mu}\left(1+\gamma^5\right)\hat L^\prime+\bar{\hat L}^\prime\left(1-\gamma^5\right)\hat\mu\right].
\label{eq:Scoupmassbasis}
\ee
The first term here is simply a coupling of the muon to $S$ with strength
\be
g_S^{\mu\mu}=-y_S\sin\theta=-\frac{y_Sy^\prime v}{\sqrt2 M}\left(1+\frac{{y^\prime}^2 v^2}{2 M^2}\right)^{-1/2}
\ee
which matches that found in the EFT in Eq.~(\ref{eq:gmumuEFT}) for $y^\prime v\ll M$ with $c_S=y_Sy^\prime$. Using the naturalness bounds of Eqs.~(\ref{eq:ysboundfroms2}) and (\ref{eq:ypboundfromh2}) leads to an upper bound on this coupling
\be
g_S^{\mu\mu}\lesssim\frac{16\pi^2 m_Sv^2}{\sqrt2 M^3}\simeq5\times10^{-3}\left(\frac{m_S}{0.1~\rm GeV}\right)\left(\frac{500~\rm GeV}{M}\right)^3.
\label{eq:gSlimUV}
\ee
Comparing this to Eq.~(\ref{eq:gSlimEFT}), we see, depending on the value of the EFT cutoff $M$, the naturalness bound in the UV complete theory can be more or less constraining than in the EFT. For the 500 GeV cutoff in Fig.~\ref{fig:sll}, the limit we have just derived from the renormalizable completion is no stronger than that from the EFT above.

\begin{figure}[tbp]
\centering
\includegraphics[width=0.5\textwidth]{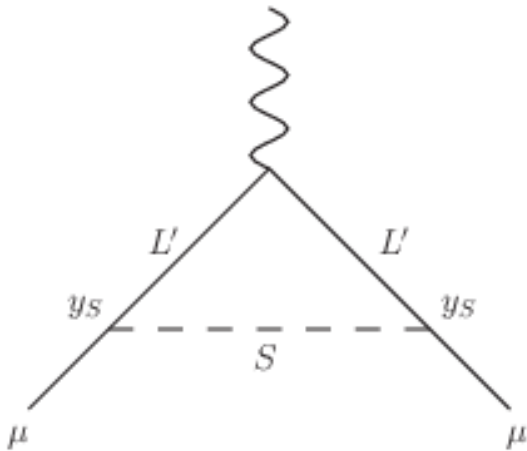}
\caption{Additional contribution to $(g-2)_\mu$ in the UV complete theory of Eq.~(\ref{eq:UVmuonlag}) from the coupling of the muon to a heavy vector-like lepton and the scalar in Eq.~(\ref{eq:Scoupmassbasis}). This contribution is generically smaller than that involving virtual muons that also appears in the EFT (see the diagram on the left of Fig.~\ref{fig:g-2}).}
\label{fig:g-2_2}
\end{figure}

The second term in Eq.~(\ref{eq:Scoupmassbasis}) describes a coupling of the muon to the heavy lepton that also gives a contribution to $\Delta a_\mu$ as shown in Fig.~\ref{fig:g-2_2}.
For $M\gg m_S,m_\mu,y^\prime v$ this is
\be
\Delta a_\mu\Big|_{L^\prime}\simeq\frac{y_S^2}{96\pi^2}\frac{m_\mu^2}{M^2}\lesssim \frac{m_\mu^2m_S^2}{6M^4}\simeq3\times10^{-16}\left(\frac{m_S}{0.1~\rm GeV}\right)^2\left(\frac{500~\rm GeV}{M}\right)^4,
\ee
where the inequality comes from the naturalness limit on $y_S$ in Eq.~(\ref{eq:ysboundfroms2}). We see, therefore, that additional contributions to $\Delta a_\mu$ from a UV completion are negligible compared to those captured in the EFT in a natural theory.

In this model, there is an additional constraint from electroweak precision tests. This comes from the fact that the right-handed muon is an admixture of $\mu_R$ and $L_R^\prime$ which have different electroweak quantum numbers. In particular, this shifts the coupling of the right-handed muon to the $Z$, $g_R$, by an amount proportional to the square of the mixing angle~\cite{delAguila:2008pw,Freitas:2014pua},
\be
\delta g_R=\sin^2\theta\left(g_L^{\rm SM}-g_R^{\rm SM}\right)\simeq\frac{{y^\prime}^2v^2}{2M^2}\left(g_L^{\rm SM}-g_R^{\rm SM}\right),
\ee
where $g_{L,R}^{\rm SM}$ are the SM values of the couplings of the left- and right-handed leptons to the $Z$. The limit on this shift from precision measurements on the $Z$ pole~\cite{Patrignani:2016xqp},
\be
\frac{y^\prime v}{M} \lesssim 0.05
\label{eq:ewprecisionlimit}
\ee
can be combined with the naturalness limit on $y_S$~(\ref{eq:ysboundfroms2}) to set an upper limit on $g_S^{\mu\mu}$. We show this limit in Fig.~\ref{fig:sll} as a brown line. We note that this simple limit rules out most of the region that can explain the $(g-2)_\mu$ discrepancy. However, we stress that this is a model-dependent limit that can be lessened or is absent in other UV completions, e.g.\ a theory with additional vector-like leptons that have the quantum numbers of right-handed leptons~\cite{Chen:2015vqy} or UV completions involving scalars instead of fermions~\cite{Batell:2016ove}.

\subsubsection{Bounds on $g_S^{\mu\mu}$}
\begin{figure}[tbp]
\centering
\includegraphics[width=\textwidth]{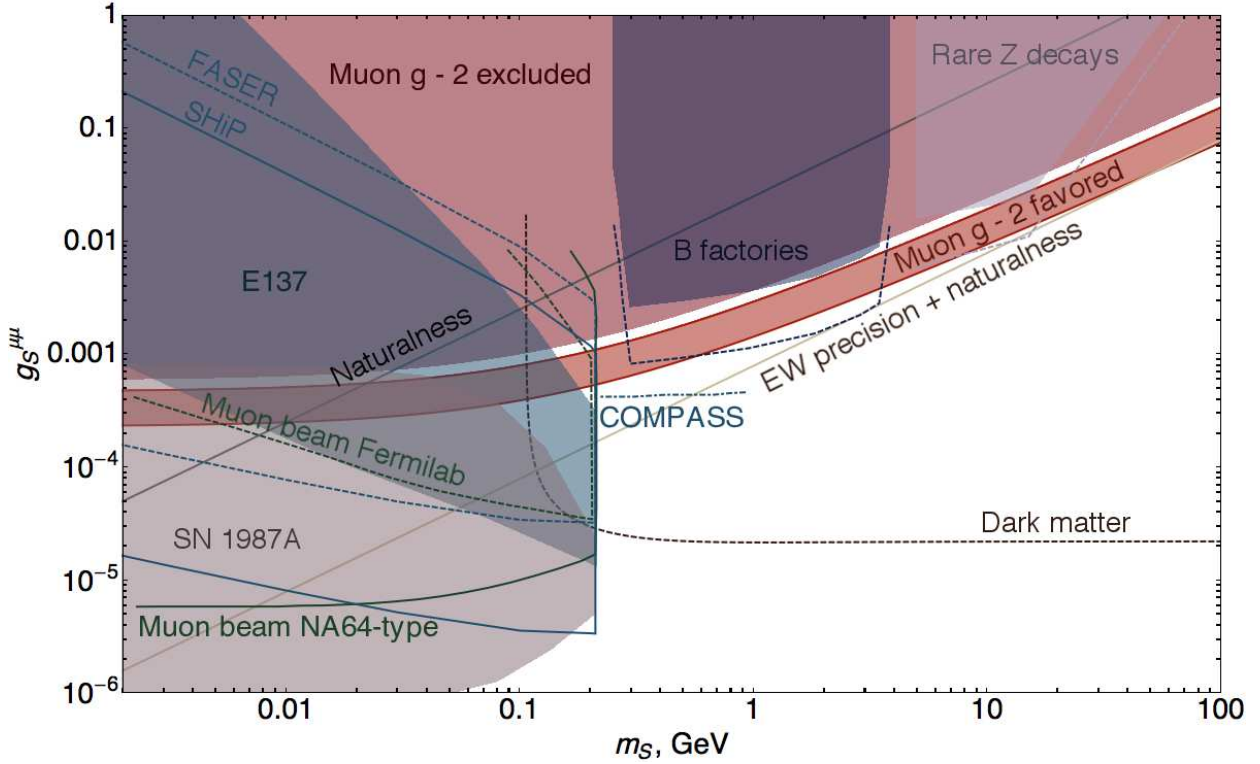}
\caption{\label{fig:sll} 
Constraints on a light scalar coupling to muons in the $m_S - g_S^{\mu\mu}$ plane. 
The orange band indicates the region of parameter space where the current $(g-2)_\mu$ discrepancy~\cite{Roberts:2010cj,Hagiwara:2011af} is below $2 \sigma$. 
The red shaded region above this band is excluded since here the $(g-2)_\mu$ discrepancy is larger than $5\sigma$. Also shown are limits from Supernova 1987A (gray shaded)~\cite{Dobrich:2015jyk,Dolan:2017osp}, SLAC beam dump E137 (blue shaded)~\cite{Bjorken:1988as,Dobrich:2015jyk,Dolan:2017osp}, BaBar (purple shaded)~\cite{TheBABAR:2016rlg}, and CMS (light purple shaded)~\cite{Sirunyan:2018nnz}. We furthermore indicate the projected sensitivity of several proposed experiments and/or analyses, including COMPASS (blue dot-dashed line)~\cite{Essig:2010gu,Abbon:2007pq}, SHiP (blue solid line)~\cite{Alekhin:2015byh}, FASER (blue dashed line)~\cite{Feng:2017uoz,Feng:2017vli}, NA64-type muon beam fixed target (green solid line)~\cite{Chen:2017awl}, Fermilab muon beam fixed target (green dashed)~\cite{Chen:2017awl}, Belle-II (purple dashed line)~\cite{Abe:2010gxa}, and HL-LHC (light purple dashed line). Assuming a coupling of the scalar to dark matter, the black dashed line 
indicates where the annihilation rate to muons is equal to the canonical thermal relic value, $\langle \sigma v \rangle = 3 \times 10^{-26}$ cm$^3$ s$^{-1}$ for $m_\chi = (1/2)(m_\mu+m_S)$ and $y_\chi < 4 \pi$.
Finally, the region below the dark brown solid line (light brown solid line) is natural according to the EFT criterion (renormalizable model criterion including electroweak precision) presented in Eq.~(\ref{eq:gSlimEFT}) (Eq.~(\ref{eq:ysboundfroms2},\ref{eq:ewprecisionlimit})). A 500 GeV cutoff scale is assumed.
}
\end{figure}

We now consider generic bounds on a scalar coupled to muons that come from beam dumps, colliders, and astrophysical observations. To study these, we first need to understand the decay channels of the scalar. For a scalar above muon threshold, its width is dominated by decays to $\mu^+\mu^-$ with a rate
\be
\Gamma_{S\to\mu^+\mu^-}=\frac{{g_S^{\mu\mu}}^2}{8\pi}m_S\left(1-\frac{4m_\mu^2}{m_S^2}\right)^{3/2}.
\ee
The $S \to \mu^+ \mu^-$ decay is generally prompt in our parameter space of interest, when kinematically allowed. In addition to the coupling to muons it is important to consider the coupling of the scalar to photons that arises due to a muon loop (the two-loop coupling to electrons is negligible). The relevant part of the effective Lagrangian containing this interaction is
\be
{\cal L}_{\rm eff}\supset \frac{g_S^{\mu\mu}\alpha}{6\pi m_\mu}F_{1/2}\left(\frac{4m_\mu^2}{m_S^2}\right)SF^{\mu\nu}F_{\mu\nu},
\label{eq:L2gamma}
\ee
where
\be
F_{1/2}\left(\tau\right)=\frac{3\tau}{2}\left[1+\left(1-\tau\right)\left(\sin^{-1}\frac{1}{\sqrt\tau}\right)^2\right].
\ee
For $m_S\ll m_\mu$, $F_{1/2}(4m_\mu^2/m_S^2)\to1$. This interaction gives a rate for $S\to\gamma\gamma$ of
\be
\Gamma_{S\to\gamma\gamma}=\frac{\alpha^2\left(g_S^{\mu\mu}\right)^2m_S^3}{144\pi^3m_\mu^2}\left| F_{1/2}\left(\frac{4m_\mu^2}{m_S^2}\right)\right|^2.
\ee
When the scalar mass is below the muon threshold, its loop-induced decay to photons can be quite slow, enabling it to be long-lived. In addition to mediating light scalar decay, the two-photon coupling can allow for $S$ to be produced in electron beam dumps as well as in supernovae. In Refs.~\cite{Dobrich:2015jyk,Dolan:2017osp}, the effects of a scalar coupled to photons through 
dimension-five operators were studied.
The lack of observation of a signal at the electron beam dump experiment E137~\cite{Bjorken:1988as} as well as the requirement that scalar production not lead to excessive cooling of supernova 1987A lead to limits on the strength of this operator.
Using the expression for the coefficient in Eq.~(\ref{eq:L2gamma}), we translate these limits on the strength of the $SF^{\mu\nu}F_{\mu\nu}$ operator into limits on $g_S^{\mu\mu}$, which we show in Fig.~\ref{fig:sll}.
Note that these limits do not apply to $m_S>2m_\mu$ since in this region, the scalar rapidly decays to $\mu^+\mu^-$.

Additionally, there are proposals to search for light scalars produced at the SHiP experiment~\cite{Alekhin:2015byh}, a proton beam dump at the CERN SPS, as well as at FASER, which is a proposal to look for particles produced at the LHC in the extreme forward direction~\cite{Feng:2017uoz,Feng:2017vli}. In both cases, we estimate the reach for muon-coupled scalars by considering production through the decay of charged kaons produced in the collisions, $K^+\to\mu^+\nu S$. For SHiP, we take estimates of the number of kaons and the energy of their decay products from~\cite{Alekhin:2015byh}. For FASER, we follow the procedure of~\cite{Feng:2017vli} and simulate forward kaon production using EPOS-LHC~\cite{Pierog:2013ria} within CRMC~\cite{crmc}. The scalars produced from kaons can then travel to the detectors where their decays can be seen. We show the regions of parameter space that can be probed at these experiments in Fig.~\ref{fig:sll}.

We have also shown in Fig.~\ref{fig:sll} the estimated reach in the coupling from production at proposed muon beam dumps estimated in Ref.~\cite{Chen:2017awl} as well as by a proposed analysis of data from the COMPASS muon beam dump~\cite{Abbon:2007pq,Essig:2010gu}.

Scalars coupled to muons can also be produced in high energy collisions in association with muons. The BaBar experiment performed a search for new vectors that couple to muons through the process $e^+e^-\to\mu^+\mu^-Z^\prime$, $Z^\prime\to\mu^+\mu^-$~\cite{TheBABAR:2016rlg}, finding no evidence for such a particle. We recast their search to find the region of parameter space ruled out for the case of a scalar, which we show in Fig.~\ref{fig:sll} as a dark purple, shaded region. The dashed line below this region shows the possible reach that the future Belle-II experiment will have given its factor of 100 increase in the amount of data compared to BaBar, assuming dominance of statistical errors. This covers a large part of parameter space that can explain the $(g-2)_\mu$ discrepancy.

The scalar $S$ can also be produced in decays of the $Z$ boson, through $Z\to\mu^+\mu^-S$ which would lead to a $Z\to4\mu$ signal. This decay mode has been measured by both ATLAS~\cite{Aad:2014wra} and CMS~\cite{Sirunyan:2018nnz}, and in the latter an explicit search for a $L_\mu - L_\tau$ gauge boson is performed. We interpret the CMS result in the context of a muon-coupled scalar using MadGraph 5~\cite{Alwall:2014hca}, deriving an upper bound on the coupling shown as the light purple shaded region in Fig.~\ref{fig:sll}. The dashed line below this region shows an estimate of the potential sensitivity in this mode that could be achieved with $3~{\rm ab}^{-1}$ of integrated luminosity at 13 TeV, assuming the same experimental cuts. In particular, this scaling assumes that high-luminosity LHC triggers will efficiently be able to capture $4\mu$ events with the leading two muons having $p_T$ above 20 and 10 GeV, respectively.

We see that for $m_S < 2 m_\mu$, E137 and SN 1987A provide strong bounds. Not only do existing experiments rule out a muon-specific scalar as an explanation of the measured muon anomalous magnetic moment, but they also cover much of the parameter space that will be probed by proposed experiments in this model. At higher masses, existing bounds do not constrain the muon-specific scalar as an explanation of $(g - 2)_{\mu}$, but with more integrated luminosity much of the relevant region of Fig.~\ref{fig:sll} will be covered by Belle-II~\cite{Abe:2010gxa}, and HL-LHC.

Finally, thus far we have assumed in this section that the $S$ interacts only with the muon at tree level. As discussed above, we can add a coupling to a DM particle, which we take to be a Dirac fermion $\chi$. Then, it is possible that the $S\bar{\mu}\mu$ coupling is connected to the DM abundance. If $m_\chi > m_S$, the secluded annihilation channel $\bar{\chi} \chi \to S S$ can annihilate away the $\chi$ population independently of the $S$ coupling to the SM. On the other hand, for $m_S > m_\chi > m_\mu$, the DM undergoes annihilation to muons. The annihilation cross section is given by an expression similar to Eq.~\eqref{eq:sigmav} with an extra factor of $1/3$ for color and appropriate kinematic factors if the muon mass is not negligible. With this in mind, we choose the benchmark mass $m_{\chi} = \frac{1}{2} \left( m_\mu + m_S \right)$. Then, the curve labeled ``Dark matter'' in Fig.~\ref{fig:sll} represents the minimum $S \bar{\mu} \mu$ coupling needed to achieve the observed DM relic density with a standard thermal cosmology, i.e. assuming that the scalar coupling to DM is $y_\chi \sim 4\pi$.\footnote{One may also consider a naturalness criterion for $y_\chi$, but it is generally relaxed with respect to the $g_S^{\mu\mu}$ naturalness bound by a factor $(m_{\rm DM}/M)^2$, where $m_{\rm DM}$ is either $m_\chi$ or some UV scale in the dark sector (which may be significantly smaller than $M$).}
Evidently it is challenging to robustly probe thermal dark matter that annihilates to muons in this scenario.

\section{Conclusions}
\label{sec:concl}

New light scalars are ubiquitous in BSM physics. The most commonly used framework for avoiding flavor constraints is to assume that any new scalar has couplings to the SM fermions that are proportional to the Yukawa couplings. Theories that satisfy the resulting MFV paradigm are safe from FCNCs, but represent only a subset of possible models with underlying flavor patterns that evade flavor bounds. In this work, by contrast we have considered an alternative class of \emph{flavor-specific} scalar models in which a new scalar couples dominantly to the first or second generation. At the price of assuming alignment between the flavor symmetry broken by a single fermion Yukawa and that broken by the coupling of a new scalar, one obtains symmetry breaking patterns which naturally suppress FCNCs.

By treating the scalar couplings as flavor symmetry spurions, we have not only explicitly demonstrated that they can be naturally small, but also parametrized the eventual flavor violation in terms of these couplings. Generally all FCNCs are suppressed by small Yukawas in our approach. While we have focused on phenomenology, it would be interesting to consider the realization of our underlying symmetry structure from a UV perspective, along recent avenues of investigation~\cite{Knapen:2015hia,Altmannshofer:2017uvs}. Nevertheless, we have gone beyond the use of simple effective operators to describe the interaction between a new scalar and the SM, examining possible renormalizable models and their implications for naturalness.

The new scalars which we have studied are useful in many contexts. We have considered a sampling of flavor-specific scalar models as an application of our framework. In particular, we have reviewed the potential constraints on a scalar which mediates interactions between the up quark and DM. Between direct detection, indirect detection, and neutron EDM searches, it is challenging to choose couplings of the new mediator to the up quark and DM such that a signature of thermal DM annihilating to up quarks with a mass below the electroweak scale would not yet have been observed. We have also examined a muon-specific scalar, which offers a potential resolution of the discrepancy between the observed and measured anomalous magnetic moment of the muon. If such a scalar weighs less than $2 m_\mu$, existing beam dump and supernova observations sharply bound its muon coupling, challenging a possible resolution of the discrepancy. Unlike its spin-1 counterpart, however, a muon-specific scalar at relatively large mass does not seem to be limited as strongly by existing constraints, such as those from B-factories and the LHC.

Models with new spin-0 particles represent a unique class of new physics theories, and in considering general flavor symmetries that can set the coupling structure, we have put flavor-specific scalars on firm theoretical ground. We hope that our results prove useful in phenomenological constructions of BSM theories with additional scalars.

\acknowledgments

We thank B.~Carlson, Z. Chacko, C.-Y.~Chen, T.~M.~Hong, A.~Hook, M.~Pospelov, D.~Robinson, J.~Wells, and C.~Zhang for useful discussions. The work of BB, AI and DM is supported in part by the U.S.~Department of Energy under grant No.~DE-SC0015634, and in part by PITT PACC. The work of AF is supported in part by the National Science Foundation under grant No.~PHY-1519175.

\input{lightscalar.bbl}

\end{document}

%% file: lightscalar.bbl
%

%% file: lightscalar.bbl
\begin{thebibliography}{89}%
\makeatletter
\providecommand \@ifxundefined [1]{%
 \@ifx{#1\undefined}
}%
\providecommand \@ifnum [1]{%
 \ifnum #1\expandafter \@firstoftwo
 \else \expandafter \@secondoftwo
 \fi
}%
\providecommand \@ifx [1]{%
 \ifx #1\expandafter \@firstoftwo
 \else \expandafter \@secondoftwo
 \fi
}%
\providecommand \natexlab [1]{#1}%
\providecommand \enquote  [1]{``#1''}%
\providecommand \bibnamefont  [1]{#1}%
\providecommand \bibfnamefont [1]{#1}%
\providecommand \citenamefont [1]{#1}%
\providecommand \href@noop [0]{\@secondoftwo}%
\providecommand \href [0]{\begingroup \@sanitize@url \@href}%
\providecommand \@href[1]{\@@startlink{#1}\@@href}%
\providecommand \@@href[1]{\endgroup#1\@@endlink}%
\providecommand \@sanitize@url [0]{\catcode `\\12\catcode `\$12\catcode
  `\&12\catcode `\#12\catcode `\^12\catcode `\_12\catcode `\%12\relax}%
\providecommand \@@startlink[1]{}%
\providecommand \@@endlink[0]{}%
\providecommand \url  [0]{\begingroup\@sanitize@url \@url }%
\providecommand \@url [1]{\endgroup\@href {#1}{\urlprefix }}%
\providecommand \urlprefix  [0]{URL }%
\providecommand \Eprint [0]{\href }%
\providecommand \doibase [0]{http://dx.doi.org/}%
\providecommand \selectlanguage [0]{\@gobble}%
\providecommand \bibinfo  [0]{\@secondoftwo}%
\providecommand \bibfield  [0]{\@secondoftwo}%
\providecommand \translation [1]{[#1]}%
\providecommand \BibitemOpen [0]{}%
\providecommand \bibitemStop [0]{}%
\providecommand \bibitemNoStop [0]{.\EOS\space}%
\providecommand \EOS [0]{\spacefactor3000\relax}%
\providecommand \BibitemShut  [1]{\csname bibitem#1\endcsname}%
\let\auto@bib@innerbib\@empty
\bibitem [{\citenamefont {Peccei}\ and\ \citenamefont
  {Quinn}(1977{\natexlab{a}})}]{Peccei:1977np}%
  \BibitemOpen
  \bibfield  {author} {\bibinfo {author} {\bibfnamefont {R.~D.}\ \bibnamefont
  {Peccei}}\ and\ \bibinfo {author} {\bibfnamefont {H.~R.}\ \bibnamefont
  {Quinn}},\ }\href {\doibase 10.1007/BF02730110} {\bibfield  {journal}
  {\bibinfo  {journal} {Nuovo Cim.}\ }\textbf {\bibinfo {volume} {A41}},\
  \bibinfo {pages} {309} (\bibinfo {year} {1977}{\natexlab{a}})}\BibitemShut
  {NoStop}%
\bibitem [{\citenamefont {Peccei}\ and\ \citenamefont
  {Quinn}(1977{\natexlab{b}})}]{Peccei:1977hh}%
  \BibitemOpen
  \bibfield  {author} {\bibinfo {author} {\bibfnamefont {R.~D.}\ \bibnamefont
  {Peccei}}\ and\ \bibinfo {author} {\bibfnamefont {H.~R.}\ \bibnamefont
  {Quinn}},\ }\href {\doibase 10.1103/PhysRevLett.38.1440} {\bibfield
  {journal} {\bibinfo  {journal} {Phys. Rev. Lett.}\ }\textbf {\bibinfo
  {volume} {38}},\ \bibinfo {pages} {1440} (\bibinfo {year}
  {1977}{\natexlab{b}})}\BibitemShut {NoStop}%
\bibitem [{\citenamefont {Weinberg}(1978)}]{Weinberg:1977ma}%
  \BibitemOpen
  \bibfield  {author} {\bibinfo {author} {\bibfnamefont {S.}~\bibnamefont
  {Weinberg}},\ }\href {\doibase 10.1103/PhysRevLett.40.223} {\bibfield
  {journal} {\bibinfo  {journal} {Phys. Rev. Lett.}\ }\textbf {\bibinfo
  {volume} {40}},\ \bibinfo {pages} {223} (\bibinfo {year} {1978})}\BibitemShut
  {NoStop}%
\bibitem [{\citenamefont {Wilczek}(1978)}]{Wilczek:1977pj}%
  \BibitemOpen
  \bibfield  {author} {\bibinfo {author} {\bibfnamefont {F.}~\bibnamefont
  {Wilczek}},\ }\href {\doibase 10.1103/PhysRevLett.40.279} {\bibfield
  {journal} {\bibinfo  {journal} {Phys. Rev. Lett.}\ }\textbf {\bibinfo
  {volume} {40}},\ \bibinfo {pages} {279} (\bibinfo {year} {1978})}\BibitemShut
  {NoStop}%
\bibitem [{\citenamefont {Silveira}\ and\ \citenamefont
  {Zee}(1985)}]{Silveira:1985rk}%
  \BibitemOpen
  \bibfield  {author} {\bibinfo {author} {\bibfnamefont {V.}~\bibnamefont
  {Silveira}}\ and\ \bibinfo {author} {\bibfnamefont {A.}~\bibnamefont {Zee}},\
  }\href {\doibase 10.1016/0370-2693(85)90624-0} {\bibfield  {journal}
  {\bibinfo  {journal} {Phys. Lett.}\ }\textbf {\bibinfo {volume} {161B}},\
  \bibinfo {pages} {136} (\bibinfo {year} {1985})}\BibitemShut {NoStop}%
\bibitem [{\citenamefont {McDonald}(1994)}]{McDonald:1993ex}%
  \BibitemOpen
  \bibfield  {author} {\bibinfo {author} {\bibfnamefont {J.}~\bibnamefont
  {McDonald}},\ }\href {\doibase 10.1103/PhysRevD.50.3637} {\bibfield
  {journal} {\bibinfo  {journal} {Phys. Rev.}\ }\textbf {\bibinfo {volume}
  {D50}},\ \bibinfo {pages} {3637} (\bibinfo {year} {1994})},\ \Eprint
  {http://arxiv.org/abs/hep-ph/0702143} {arXiv:hep-ph/0702143 [HEP-PH]}
  \BibitemShut {NoStop}%
\bibitem [{\citenamefont {Burgess}\ \emph {et~al.}(2001)\citenamefont
  {Burgess}, \citenamefont {Pospelov},\ and\ \citenamefont {ter
  Veldhuis}}]{Burgess:2000yq}%
  \BibitemOpen
  \bibfield  {author} {\bibinfo {author} {\bibfnamefont {C.~P.}\ \bibnamefont
  {Burgess}}, \bibinfo {author} {\bibfnamefont {M.}~\bibnamefont {Pospelov}}, \
  and\ \bibinfo {author} {\bibfnamefont {T.}~\bibnamefont {ter Veldhuis}},\
  }\href {\doibase 10.1016/S0550-3213(01)00513-2} {\bibfield  {journal}
  {\bibinfo  {journal} {Nucl. Phys.}\ }\textbf {\bibinfo {volume} {B619}},\
  \bibinfo {pages} {709} (\bibinfo {year} {2001})},\ \Eprint
  {http://arxiv.org/abs/hep-ph/0011335} {arXiv:hep-ph/0011335 [hep-ph]}
  \BibitemShut {NoStop}%
\bibitem [{\citenamefont {Boehm}\ and\ \citenamefont
  {Fayet}(2004)}]{Boehm:2003hm}%
  \BibitemOpen
  \bibfield  {author} {\bibinfo {author} {\bibfnamefont {C.}~\bibnamefont
  {Boehm}}\ and\ \bibinfo {author} {\bibfnamefont {P.}~\bibnamefont {Fayet}},\
  }\href {\doibase 10.1016/j.nuclphysb.2004.01.015} {\bibfield  {journal}
  {\bibinfo  {journal} {Nucl. Phys.}\ }\textbf {\bibinfo {volume} {B683}},\
  \bibinfo {pages} {219} (\bibinfo {year} {2004})},\ \Eprint
  {http://arxiv.org/abs/hep-ph/0305261} {arXiv:hep-ph/0305261 [hep-ph]}
  \BibitemShut {NoStop}%
\bibitem [{\citenamefont {Borodatchenkova}\ \emph {et~al.}(2006)\citenamefont
  {Borodatchenkova}, \citenamefont {Choudhury},\ and\ \citenamefont
  {Drees}}]{Borodatchenkova:2005ct}%
  \BibitemOpen
  \bibfield  {author} {\bibinfo {author} {\bibfnamefont {N.}~\bibnamefont
  {Borodatchenkova}}, \bibinfo {author} {\bibfnamefont {D.}~\bibnamefont
  {Choudhury}}, \ and\ \bibinfo {author} {\bibfnamefont {M.}~\bibnamefont
  {Drees}},\ }\href {\doibase 10.1103/PhysRevLett.96.141802} {\bibfield
  {journal} {\bibinfo  {journal} {Phys. Rev. Lett.}\ }\textbf {\bibinfo
  {volume} {96}},\ \bibinfo {pages} {141802} (\bibinfo {year} {2006})},\
  \Eprint {http://arxiv.org/abs/hep-ph/0510147} {arXiv:hep-ph/0510147 [hep-ph]}
  \BibitemShut {NoStop}%
\bibitem [{\citenamefont {Fayet}(2006)}]{Fayet:2006sp}%
  \BibitemOpen
  \bibfield  {author} {\bibinfo {author} {\bibfnamefont {P.}~\bibnamefont
  {Fayet}},\ }\href {\doibase 10.1103/PhysRevD.74.054034} {\bibfield  {journal}
  {\bibinfo  {journal} {Phys. Rev.}\ }\textbf {\bibinfo {volume} {D74}},\
  \bibinfo {pages} {054034} (\bibinfo {year} {2006})},\ \Eprint
  {http://arxiv.org/abs/hep-ph/0607318} {arXiv:hep-ph/0607318 [hep-ph]}
  \BibitemShut {NoStop}%
\bibitem [{\citenamefont {Gninenko}\ and\ \citenamefont
  {Krasnikov}(2001)}]{Gninenko:2001hx}%
  \BibitemOpen
  \bibfield  {author} {\bibinfo {author} {\bibfnamefont {S.~N.}\ \bibnamefont
  {Gninenko}}\ and\ \bibinfo {author} {\bibfnamefont {N.~V.}\ \bibnamefont
  {Krasnikov}},\ }\href {\doibase 10.1016/S0370-2693(01)00693-1} {\bibfield
  {journal} {\bibinfo  {journal} {Phys. Lett.}\ }\textbf {\bibinfo {volume}
  {B513}},\ \bibinfo {pages} {119} (\bibinfo {year} {2001})},\ \Eprint
  {http://arxiv.org/abs/hep-ph/0102222} {arXiv:hep-ph/0102222 [hep-ph]}
  \BibitemShut {NoStop}%
\bibitem [{\citenamefont {Fayet}(2007)}]{Fayet:2007ua}%
  \BibitemOpen
  \bibfield  {author} {\bibinfo {author} {\bibfnamefont {P.}~\bibnamefont
  {Fayet}},\ }\href {\doibase 10.1103/PhysRevD.75.115017} {\bibfield  {journal}
  {\bibinfo  {journal} {Phys. Rev.}\ }\textbf {\bibinfo {volume} {D75}},\
  \bibinfo {pages} {115017} (\bibinfo {year} {2007})},\ \Eprint
  {http://arxiv.org/abs/hep-ph/0702176} {arXiv:hep-ph/0702176 [HEP-PH]}
  \BibitemShut {NoStop}%
\bibitem [{\citenamefont {Pospelov}(2009)}]{Pospelov:2008zw}%
  \BibitemOpen
  \bibfield  {author} {\bibinfo {author} {\bibfnamefont {M.}~\bibnamefont
  {Pospelov}},\ }\href {\doibase 10.1103/PhysRevD.80.095002} {\bibfield
  {journal} {\bibinfo  {journal} {Phys. Rev.}\ }\textbf {\bibinfo {volume}
  {D80}},\ \bibinfo {pages} {095002} (\bibinfo {year} {2009})},\ \Eprint
  {http://arxiv.org/abs/0811.1030} {arXiv:0811.1030 [hep-ph]} \BibitemShut
  {NoStop}%
\bibitem [{\citenamefont {Davoudiasl}\ \emph {et~al.}(2012)\citenamefont
  {Davoudiasl}, \citenamefont {Lee},\ and\ \citenamefont
  {Marciano}}]{Davoudiasl:2012ig}%
  \BibitemOpen
  \bibfield  {author} {\bibinfo {author} {\bibfnamefont {H.}~\bibnamefont
  {Davoudiasl}}, \bibinfo {author} {\bibfnamefont {H.-S.}\ \bibnamefont {Lee}},
  \ and\ \bibinfo {author} {\bibfnamefont {W.~J.}\ \bibnamefont {Marciano}},\
  }\href {\doibase 10.1103/PhysRevD.86.095009} {\bibfield  {journal} {\bibinfo
  {journal} {Phys. Rev.}\ }\textbf {\bibinfo {volume} {D86}},\ \bibinfo {pages}
  {095009} (\bibinfo {year} {2012})},\ \Eprint {http://arxiv.org/abs/1208.2973}
  {arXiv:1208.2973 [hep-ph]} \BibitemShut {NoStop}%
\bibitem [{\citenamefont {Tucker-Smith}\ and\ \citenamefont
  {Yavin}(2011)}]{TuckerSmith:2010ra}%
  \BibitemOpen
  \bibfield  {author} {\bibinfo {author} {\bibfnamefont {D.}~\bibnamefont
  {Tucker-Smith}}\ and\ \bibinfo {author} {\bibfnamefont {I.}~\bibnamefont
  {Yavin}},\ }\href {\doibase 10.1103/PhysRevD.83.101702} {\bibfield  {journal}
  {\bibinfo  {journal} {Phys. Rev.}\ }\textbf {\bibinfo {volume} {D83}},\
  \bibinfo {pages} {101702} (\bibinfo {year} {2011})},\ \Eprint
  {http://arxiv.org/abs/1011.4922} {arXiv:1011.4922 [hep-ph]} \BibitemShut
  {NoStop}%
\bibitem [{\citenamefont {Batell}\ \emph {et~al.}(2011)\citenamefont {Batell},
  \citenamefont {McKeen},\ and\ \citenamefont {Pospelov}}]{Batell:2011qq}%
  \BibitemOpen
  \bibfield  {author} {\bibinfo {author} {\bibfnamefont {B.}~\bibnamefont
  {Batell}}, \bibinfo {author} {\bibfnamefont {D.}~\bibnamefont {McKeen}}, \
  and\ \bibinfo {author} {\bibfnamefont {M.}~\bibnamefont {Pospelov}},\ }\href
  {\doibase 10.1103/PhysRevLett.107.011803} {\bibfield  {journal} {\bibinfo
  {journal} {Phys. Rev. Lett.}\ }\textbf {\bibinfo {volume} {107}},\ \bibinfo
  {pages} {011803} (\bibinfo {year} {2011})},\ \Eprint
  {http://arxiv.org/abs/1103.0721} {arXiv:1103.0721 [hep-ph]} \BibitemShut
  {NoStop}%
\bibitem [{\citenamefont {Schmidt-Hoberg}\ \emph {et~al.}(2013)\citenamefont
  {Schmidt-Hoberg}, \citenamefont {Staub},\ and\ \citenamefont
  {Winkler}}]{Schmidt-Hoberg:2013hba}%
  \BibitemOpen
  \bibfield  {author} {\bibinfo {author} {\bibfnamefont {K.}~\bibnamefont
  {Schmidt-Hoberg}}, \bibinfo {author} {\bibfnamefont {F.}~\bibnamefont
  {Staub}}, \ and\ \bibinfo {author} {\bibfnamefont {M.~W.}\ \bibnamefont
  {Winkler}},\ }\href {\doibase 10.1016/j.physletb.2013.11.015} {\bibfield
  {journal} {\bibinfo  {journal} {Phys. Lett.}\ }\textbf {\bibinfo {volume}
  {B727}},\ \bibinfo {pages} {506} (\bibinfo {year} {2013})},\ \Eprint
  {http://arxiv.org/abs/1310.6752} {arXiv:1310.6752 [hep-ph]} \BibitemShut
  {NoStop}%
\bibitem [{\citenamefont {Clarke}\ \emph {et~al.}(2014)\citenamefont {Clarke},
  \citenamefont {Foot},\ and\ \citenamefont {Volkas}}]{Clarke:2013aya}%
  \BibitemOpen
  \bibfield  {author} {\bibinfo {author} {\bibfnamefont {J.~D.}\ \bibnamefont
  {Clarke}}, \bibinfo {author} {\bibfnamefont {R.}~\bibnamefont {Foot}}, \ and\
  \bibinfo {author} {\bibfnamefont {R.~R.}\ \bibnamefont {Volkas}},\ }\href
  {\doibase 10.1007/JHEP02(2014)123} {\bibfield  {journal} {\bibinfo  {journal}
  {JHEP}\ }\textbf {\bibinfo {volume} {02}},\ \bibinfo {pages} {123} (\bibinfo
  {year} {2014})},\ \Eprint {http://arxiv.org/abs/1310.8042} {arXiv:1310.8042
  [hep-ph]} \BibitemShut {NoStop}%
\bibitem [{\citenamefont {Chen}\ \emph {et~al.}(2016)\citenamefont {Chen},
  \citenamefont {Davoudiasl}, \citenamefont {Marciano},\ and\ \citenamefont
  {Zhang}}]{Chen:2015vqy}%
  \BibitemOpen
  \bibfield  {author} {\bibinfo {author} {\bibfnamefont {C.-Y.}\ \bibnamefont
  {Chen}}, \bibinfo {author} {\bibfnamefont {H.}~\bibnamefont {Davoudiasl}},
  \bibinfo {author} {\bibfnamefont {W.~J.}\ \bibnamefont {Marciano}}, \ and\
  \bibinfo {author} {\bibfnamefont {C.}~\bibnamefont {Zhang}},\ }\href
  {\doibase 10.1103/PhysRevD.93.035006} {\bibfield  {journal} {\bibinfo
  {journal} {Phys. Rev.}\ }\textbf {\bibinfo {volume} {D93}},\ \bibinfo {pages}
  {035006} (\bibinfo {year} {2016})},\ \Eprint
  {http://arxiv.org/abs/1511.04715} {arXiv:1511.04715 [hep-ph]} \BibitemShut
  {NoStop}%
\bibitem [{\citenamefont {Batell}\ \emph {et~al.}(2017)\citenamefont {Batell},
  \citenamefont {Lange}, \citenamefont {McKeen}, \citenamefont {Pospelov},\
  and\ \citenamefont {Ritz}}]{Batell:2016ove}%
  \BibitemOpen
  \bibfield  {author} {\bibinfo {author} {\bibfnamefont {B.}~\bibnamefont
  {Batell}}, \bibinfo {author} {\bibfnamefont {N.}~\bibnamefont {Lange}},
  \bibinfo {author} {\bibfnamefont {D.}~\bibnamefont {McKeen}}, \bibinfo
  {author} {\bibfnamefont {M.}~\bibnamefont {Pospelov}}, \ and\ \bibinfo
  {author} {\bibfnamefont {A.}~\bibnamefont {Ritz}},\ }\href {\doibase
  10.1103/PhysRevD.95.075003} {\bibfield  {journal} {\bibinfo  {journal} {Phys.
  Rev.}\ }\textbf {\bibinfo {volume} {D95}},\ \bibinfo {pages} {075003}
  (\bibinfo {year} {2017})},\ \Eprint {http://arxiv.org/abs/1606.04943}
  {arXiv:1606.04943 [hep-ph]} \BibitemShut {NoStop}%
\bibitem [{\citenamefont {Preskill}(1991)}]{Preskill:1990fr}%
  \BibitemOpen
  \bibfield  {author} {\bibinfo {author} {\bibfnamefont {J.}~\bibnamefont
  {Preskill}},\ }\href {\doibase 10.1016/0003-4916(91)90046-B} {\bibfield
  {journal} {\bibinfo  {journal} {Annals Phys.}\ }\textbf {\bibinfo {volume}
  {210}},\ \bibinfo {pages} {323} (\bibinfo {year} {1991})}\BibitemShut
  {NoStop}%
\bibitem [{\citenamefont {Batra}\ \emph {et~al.}(2006)\citenamefont {Batra},
  \citenamefont {Dobrescu},\ and\ \citenamefont {Spivak}}]{Batra:2005rh}%
  \BibitemOpen
  \bibfield  {author} {\bibinfo {author} {\bibfnamefont {P.}~\bibnamefont
  {Batra}}, \bibinfo {author} {\bibfnamefont {B.~A.}\ \bibnamefont {Dobrescu}},
  \ and\ \bibinfo {author} {\bibfnamefont {D.}~\bibnamefont {Spivak}},\ }\href
  {\doibase 10.1063/1.2222081} {\bibfield  {journal} {\bibinfo  {journal} {J.
  Math. Phys.}\ }\textbf {\bibinfo {volume} {47}},\ \bibinfo {pages} {082301}
  (\bibinfo {year} {2006})},\ \Eprint {http://arxiv.org/abs/hep-ph/0510181}
  {arXiv:hep-ph/0510181 [hep-ph]} \BibitemShut {NoStop}%
\bibitem [{\citenamefont {Dror}\ \emph
  {et~al.}(2017{\natexlab{a}})\citenamefont {Dror}, \citenamefont {Lasenby},\
  and\ \citenamefont {Pospelov}}]{Dror:2017ehi}%
  \BibitemOpen
  \bibfield  {author} {\bibinfo {author} {\bibfnamefont {J.~A.}\ \bibnamefont
  {Dror}}, \bibinfo {author} {\bibfnamefont {R.}~\bibnamefont {Lasenby}}, \
  and\ \bibinfo {author} {\bibfnamefont {M.}~\bibnamefont {Pospelov}},\ }\href
  {\doibase 10.1103/PhysRevLett.119.141803} {\bibfield  {journal} {\bibinfo
  {journal} {Phys. Rev. Lett.}\ }\textbf {\bibinfo {volume} {119}},\ \bibinfo
  {pages} {141803} (\bibinfo {year} {2017}{\natexlab{a}})},\ \Eprint
  {http://arxiv.org/abs/1705.06726} {arXiv:1705.06726 [hep-ph]} \BibitemShut
  {NoStop}%
\bibitem [{\citenamefont {Ismail}\ \emph {et~al.}(2017)\citenamefont {Ismail},
  \citenamefont {Katz},\ and\ \citenamefont {Racco}}]{Ismail:2017ulg}%
  \BibitemOpen
  \bibfield  {author} {\bibinfo {author} {\bibfnamefont {A.}~\bibnamefont
  {Ismail}}, \bibinfo {author} {\bibfnamefont {A.}~\bibnamefont {Katz}}, \ and\
  \bibinfo {author} {\bibfnamefont {D.}~\bibnamefont {Racco}},\ }\href
  {\doibase 10.1007/JHEP10(2017)165} {\bibfield  {journal} {\bibinfo  {journal}
  {JHEP}\ }\textbf {\bibinfo {volume} {10}},\ \bibinfo {pages} {165} (\bibinfo
  {year} {2017})},\ \Eprint {http://arxiv.org/abs/1707.00709} {arXiv:1707.00709
  [hep-ph]} \BibitemShut {NoStop}%
\bibitem [{\citenamefont {Dror}\ \emph
  {et~al.}(2017{\natexlab{b}})\citenamefont {Dror}, \citenamefont {Lasenby},\
  and\ \citenamefont {Pospelov}}]{Dror:2017nsg}%
  \BibitemOpen
  \bibfield  {author} {\bibinfo {author} {\bibfnamefont {J.~A.}\ \bibnamefont
  {Dror}}, \bibinfo {author} {\bibfnamefont {R.}~\bibnamefont {Lasenby}}, \
  and\ \bibinfo {author} {\bibfnamefont {M.}~\bibnamefont {Pospelov}},\ }\href
  {\doibase 10.1103/PhysRevD.96.075036} {\bibfield  {journal} {\bibinfo
  {journal} {Phys. Rev.}\ }\textbf {\bibinfo {volume} {D96}},\ \bibinfo {pages}
  {075036} (\bibinfo {year} {2017}{\natexlab{b}})},\ \Eprint
  {http://arxiv.org/abs/1707.01503} {arXiv:1707.01503 [hep-ph]} \BibitemShut
  {NoStop}%
\bibitem [{\citenamefont {Ismail}\ and\ \citenamefont
  {Katz}(2017)}]{Ismail:2017fgq}%
  \BibitemOpen
  \bibfield  {author} {\bibinfo {author} {\bibfnamefont {A.}~\bibnamefont
  {Ismail}}\ and\ \bibinfo {author} {\bibfnamefont {A.}~\bibnamefont {Katz}},\
  }\href@noop {} {\  (\bibinfo {year} {2017})},\ \Eprint
  {http://arxiv.org/abs/1712.01840} {arXiv:1712.01840 [hep-ph]} \BibitemShut
  {NoStop}%
\bibitem [{\citenamefont {Omura}\ \emph {et~al.}(2015)\citenamefont {Omura},
  \citenamefont {Senaha},\ and\ \citenamefont {Tobe}}]{Omura:2015nja}%
  \BibitemOpen
  \bibfield  {author} {\bibinfo {author} {\bibfnamefont {Y.}~\bibnamefont
  {Omura}}, \bibinfo {author} {\bibfnamefont {E.}~\bibnamefont {Senaha}}, \
  and\ \bibinfo {author} {\bibfnamefont {K.}~\bibnamefont {Tobe}},\ }\href
  {\doibase 10.1007/JHEP05(2015)028} {\bibfield  {journal} {\bibinfo  {journal}
  {JHEP}\ }\textbf {\bibinfo {volume} {05}},\ \bibinfo {pages} {028} (\bibinfo
  {year} {2015})},\ \Eprint {http://arxiv.org/abs/1502.07824} {arXiv:1502.07824
  [hep-ph]} \BibitemShut {NoStop}%
\bibitem [{\citenamefont {Carlson}\ and\ \citenamefont
  {Freid}(2015)}]{Carlson:2015poa}%
  \BibitemOpen
  \bibfield  {author} {\bibinfo {author} {\bibfnamefont {C.~E.}\ \bibnamefont
  {Carlson}}\ and\ \bibinfo {author} {\bibfnamefont {M.}~\bibnamefont
  {Freid}},\ }\href {\doibase 10.1103/PhysRevD.92.095024} {\bibfield  {journal}
  {\bibinfo  {journal} {Phys. Rev.}\ }\textbf {\bibinfo {volume} {D92}},\
  \bibinfo {pages} {095024} (\bibinfo {year} {2015})},\ \Eprint
  {http://arxiv.org/abs/1506.06631} {arXiv:1506.06631 [hep-ph]} \BibitemShut
  {NoStop}%
\bibitem [{\citenamefont {D'Ambrosio}\ \emph {et~al.}(2002)\citenamefont
  {D'Ambrosio}, \citenamefont {Giudice}, \citenamefont {Isidori},\ and\
  \citenamefont {Strumia}}]{DAmbrosio:2002vsn}%
  \BibitemOpen
  \bibfield  {author} {\bibinfo {author} {\bibfnamefont {G.}~\bibnamefont
  {D'Ambrosio}}, \bibinfo {author} {\bibfnamefont {G.~F.}\ \bibnamefont
  {Giudice}}, \bibinfo {author} {\bibfnamefont {G.}~\bibnamefont {Isidori}}, \
  and\ \bibinfo {author} {\bibfnamefont {A.}~\bibnamefont {Strumia}},\ }\href
  {\doibase 10.1016/S0550-3213(02)00836-2} {\bibfield  {journal} {\bibinfo
  {journal} {Nucl. Phys.}\ }\textbf {\bibinfo {volume} {B645}},\ \bibinfo
  {pages} {155} (\bibinfo {year} {2002})},\ \Eprint
  {http://arxiv.org/abs/hep-ph/0207036} {arXiv:hep-ph/0207036 [hep-ph]}
  \BibitemShut {NoStop}%
\bibitem [{Note1()}]{Note1}%
  \BibitemOpen
  \bibinfo {note} {Of course, hypercharge is also conserved. Including a global
  $U(1)_H$ factor for the Higgs, the full breaking pattern is $U(3)_Q \times
  U(3)_U \times U(3)_D \times U(1)_H \rightarrow U(1)_{B} \times
  U(1)_Y$.}\BibitemShut {Stop}%
\bibitem [{\citenamefont {Agashe}\ \emph {et~al.}(2005)\citenamefont {Agashe},
  \citenamefont {Papucci}, \citenamefont {Perez},\ and\ \citenamefont
  {Pirjol}}]{Agashe:2005hk}%
  \BibitemOpen
  \bibfield  {author} {\bibinfo {author} {\bibfnamefont {K.}~\bibnamefont
  {Agashe}}, \bibinfo {author} {\bibfnamefont {M.}~\bibnamefont {Papucci}},
  \bibinfo {author} {\bibfnamefont {G.}~\bibnamefont {Perez}}, \ and\ \bibinfo
  {author} {\bibfnamefont {D.}~\bibnamefont {Pirjol}},\ }\href@noop {} {\
  (\bibinfo {year} {2005})},\ \Eprint {http://arxiv.org/abs/hep-ph/0509117}
  {arXiv:hep-ph/0509117 [hep-ph]} \BibitemShut {NoStop}%
\bibitem [{\citenamefont {Barbieri}\ \emph
  {et~al.}(2012{\natexlab{a}})\citenamefont {Barbieri}, \citenamefont
  {Buttazzo}, \citenamefont {Sala},\ and\ \citenamefont
  {Straub}}]{Barbieri:2012uh}%
  \BibitemOpen
  \bibfield  {author} {\bibinfo {author} {\bibfnamefont {R.}~\bibnamefont
  {Barbieri}}, \bibinfo {author} {\bibfnamefont {D.}~\bibnamefont {Buttazzo}},
  \bibinfo {author} {\bibfnamefont {F.}~\bibnamefont {Sala}}, \ and\ \bibinfo
  {author} {\bibfnamefont {D.~M.}\ \bibnamefont {Straub}},\ }\href {\doibase
  10.1007/JHEP07(2012)181} {\bibfield  {journal} {\bibinfo  {journal} {JHEP}\
  }\textbf {\bibinfo {volume} {07}},\ \bibinfo {pages} {181} (\bibinfo {year}
  {2012}{\natexlab{a}})},\ \Eprint {http://arxiv.org/abs/1203.4218}
  {arXiv:1203.4218 [hep-ph]} \BibitemShut {NoStop}%
\bibitem [{\citenamefont {Barbieri}\ \emph
  {et~al.}(2012{\natexlab{b}})\citenamefont {Barbieri}, \citenamefont
  {Buttazzo}, \citenamefont {Sala},\ and\ \citenamefont
  {Straub}}]{Barbieri:2012bh}%
  \BibitemOpen
  \bibfield  {author} {\bibinfo {author} {\bibfnamefont {R.}~\bibnamefont
  {Barbieri}}, \bibinfo {author} {\bibfnamefont {D.}~\bibnamefont {Buttazzo}},
  \bibinfo {author} {\bibfnamefont {F.}~\bibnamefont {Sala}}, \ and\ \bibinfo
  {author} {\bibfnamefont {D.~M.}\ \bibnamefont {Straub}},\ }\href {\doibase
  10.1007/JHEP10(2012)040} {\bibfield  {journal} {\bibinfo  {journal} {JHEP}\
  }\textbf {\bibinfo {volume} {10}},\ \bibinfo {pages} {040} (\bibinfo {year}
  {2012}{\natexlab{b}})},\ \Eprint {http://arxiv.org/abs/1206.1327}
  {arXiv:1206.1327 [hep-ph]} \BibitemShut {NoStop}%
\bibitem [{\citenamefont {Knapen}\ and\ \citenamefont
  {Robinson}(2015)}]{Knapen:2015hia}%
  \BibitemOpen
  \bibfield  {author} {\bibinfo {author} {\bibfnamefont {S.}~\bibnamefont
  {Knapen}}\ and\ \bibinfo {author} {\bibfnamefont {D.~J.}\ \bibnamefont
  {Robinson}},\ }\href {\doibase 10.1103/PhysRevLett.115.161803} {\bibfield
  {journal} {\bibinfo  {journal} {Phys. Rev. Lett.}\ }\textbf {\bibinfo
  {volume} {115}},\ \bibinfo {pages} {161803} (\bibinfo {year} {2015})},\
  \Eprint {http://arxiv.org/abs/1507.00009} {arXiv:1507.00009 [hep-ph]}
  \BibitemShut {NoStop}%
\bibitem [{\citenamefont {Altmannshofer}\ \emph {et~al.}(2018)\citenamefont
  {Altmannshofer}, \citenamefont {Gori}, \citenamefont {Robinson},\ and\
  \citenamefont {Tuckler}}]{Altmannshofer:2017uvs}%
  \BibitemOpen
  \bibfield  {author} {\bibinfo {author} {\bibfnamefont {W.}~\bibnamefont
  {Altmannshofer}}, \bibinfo {author} {\bibfnamefont {S.}~\bibnamefont {Gori}},
  \bibinfo {author} {\bibfnamefont {D.~J.}\ \bibnamefont {Robinson}}, \ and\
  \bibinfo {author} {\bibfnamefont {D.}~\bibnamefont {Tuckler}},\ }\href
  {\doibase 10.1007/JHEP03(2018)129} {\bibfield  {journal} {\bibinfo  {journal}
  {JHEP}\ }\textbf {\bibinfo {volume} {03}},\ \bibinfo {pages} {129} (\bibinfo
  {year} {2018})},\ \Eprint {http://arxiv.org/abs/1712.01847} {arXiv:1712.01847
  [hep-ph]} \BibitemShut {NoStop}%
\bibitem [{Note2()}]{Note2}%
  \BibitemOpen
  \bibinfo {note} {A similar constraint could be placed using the Higgs mass
  correction, but it would be weaker for $m_S < m_h$.}\BibitemShut {Stop}%
\bibitem [{Note3()}]{Note3}%
  \BibitemOpen
  \bibinfo {note} {We consider only couplings of $S$ to a single quark in this
  work. If $S$ couples to both up-type and down-type quarks, the flavor physics
  will be similar if the $c_S$ of Eq.\protect \textup {\hbox {\mathsurround \z@
  \protect \normalfont (\ignorespaces \ref {eq:bsmlag}\unskip \@@italiccorr )}}
  and its down-type analog are both diagonal in the respective quark mass
  bases. In this case only, the spurion arguments that follow still
  hold.}\BibitemShut {Stop}%
\bibitem [{\citenamefont {Isidori}\ \emph {et~al.}(2010)\citenamefont
  {Isidori}, \citenamefont {Nir},\ and\ \citenamefont
  {Perez}}]{Isidori:2010kg}%
  \BibitemOpen
  \bibfield  {author} {\bibinfo {author} {\bibfnamefont {G.}~\bibnamefont
  {Isidori}}, \bibinfo {author} {\bibfnamefont {Y.}~\bibnamefont {Nir}}, \ and\
  \bibinfo {author} {\bibfnamefont {G.}~\bibnamefont {Perez}},\ }\href
  {\doibase 10.1146/annurev.nucl.012809.104534} {\bibfield  {journal} {\bibinfo
   {journal} {Ann. Rev. Nucl. Part. Sci.}\ }\textbf {\bibinfo {volume} {60}},\
  \bibinfo {pages} {355} (\bibinfo {year} {2010})},\ \Eprint
  {http://arxiv.org/abs/1002.0900} {arXiv:1002.0900 [hep-ph]} \BibitemShut
  {NoStop}%
\bibitem [{\citenamefont {Blankenburg}\ \emph {et~al.}(2012)\citenamefont
  {Blankenburg}, \citenamefont {Ellis},\ and\ \citenamefont
  {Isidori}}]{Blankenburg:2012ex}%
  \BibitemOpen
  \bibfield  {author} {\bibinfo {author} {\bibfnamefont {G.}~\bibnamefont
  {Blankenburg}}, \bibinfo {author} {\bibfnamefont {J.}~\bibnamefont {Ellis}},
  \ and\ \bibinfo {author} {\bibfnamefont {G.}~\bibnamefont {Isidori}},\ }\href
  {\doibase 10.1016/j.physletb.2012.05.007} {\bibfield  {journal} {\bibinfo
  {journal} {Phys. Lett.}\ }\textbf {\bibinfo {volume} {B712}},\ \bibinfo
  {pages} {386} (\bibinfo {year} {2012})},\ \Eprint
  {http://arxiv.org/abs/1202.5704} {arXiv:1202.5704 [hep-ph]} \BibitemShut
  {NoStop}%
\bibitem [{Note4()}]{Note4}%
  \BibitemOpen
  \bibinfo {note} {We could equally well have chosen the new vector-like quark
  to have the same charge as $U_R$, which would not significantly affect the
  influence of electroweak precision constraints.}\BibitemShut {Stop}%
\bibitem [{\citenamefont {Belanger}\ \emph {et~al.}(2014)\citenamefont
  {Belanger}, \citenamefont {Boudjema}, \citenamefont {Pukhov},\ and\
  \citenamefont {Semenov}}]{Belanger:2013oya}%
  \BibitemOpen
  \bibfield  {author} {\bibinfo {author} {\bibfnamefont {G.}~\bibnamefont
  {Belanger}}, \bibinfo {author} {\bibfnamefont {F.}~\bibnamefont {Boudjema}},
  \bibinfo {author} {\bibfnamefont {A.}~\bibnamefont {Pukhov}}, \ and\ \bibinfo
  {author} {\bibfnamefont {A.}~\bibnamefont {Semenov}},\ }\href {\doibase
  10.1016/j.cpc.2013.10.016} {\bibfield  {journal} {\bibinfo  {journal}
  {Comput. Phys. Commun.}\ }\textbf {\bibinfo {volume} {185}},\ \bibinfo
  {pages} {960} (\bibinfo {year} {2014})},\ \Eprint
  {http://arxiv.org/abs/1305.0237} {arXiv:1305.0237 [hep-ph]} \BibitemShut
  {NoStop}%
\bibitem [{\citenamefont {Pendlebury}\ \emph {et~al.}(2015)\citenamefont
  {Pendlebury} \emph {et~al.}}]{Afach:2015sja}%
  \BibitemOpen
  \bibfield  {author} {\bibinfo {author} {\bibfnamefont {J.~M.}\ \bibnamefont
  {Pendlebury}} \emph {et~al.},\ }\href {\doibase 10.1103/PhysRevD.92.092003}
  {\bibfield  {journal} {\bibinfo  {journal} {Phys. Rev.}\ }\textbf {\bibinfo
  {volume} {D92}},\ \bibinfo {pages} {092003} (\bibinfo {year} {2015})},\
  \Eprint {http://arxiv.org/abs/1509.04411} {arXiv:1509.04411 [hep-ex]}
  \BibitemShut {NoStop}%
\bibitem [{\citenamefont {Crewther}\ \emph {et~al.}(1979)\citenamefont
  {Crewther}, \citenamefont {Di~Vecchia}, \citenamefont {Veneziano},\ and\
  \citenamefont {Witten}}]{Crewther:1979pi}%
  \BibitemOpen
  \bibfield  {author} {\bibinfo {author} {\bibfnamefont {R.~J.}\ \bibnamefont
  {Crewther}}, \bibinfo {author} {\bibfnamefont {P.}~\bibnamefont
  {Di~Vecchia}}, \bibinfo {author} {\bibfnamefont {G.}~\bibnamefont
  {Veneziano}}, \ and\ \bibinfo {author} {\bibfnamefont {E.}~\bibnamefont
  {Witten}},\ }\href {\doibase 10.1016/0370-2693(80)91025-4,
  10.1016/0370-2693(79)90128-X} {\bibfield  {journal} {\bibinfo  {journal}
  {Phys. Lett.}\ }\textbf {\bibinfo {volume} {88B}},\ \bibinfo {pages} {123}
  (\bibinfo {year} {1979})},\ \bibinfo {note} {[Erratum: Phys.
  Lett.91B,487(1980)]}\BibitemShut {NoStop}%
\bibitem [{\citenamefont {Baron}\ \emph {et~al.}(2014)\citenamefont {Baron}
  \emph {et~al.}}]{Baron:2013eja}%
  \BibitemOpen
  \bibfield  {author} {\bibinfo {author} {\bibfnamefont {J.}~\bibnamefont
  {Baron}} \emph {et~al.} (\bibinfo {collaboration} {ACME}),\ }\href {\doibase
  10.1126/science.1248213} {\bibfield  {journal} {\bibinfo  {journal}
  {Science}\ }\textbf {\bibinfo {volume} {343}},\ \bibinfo {pages} {269}
  (\bibinfo {year} {2014})},\ \Eprint {http://arxiv.org/abs/1310.7534}
  {arXiv:1310.7534 [physics.atom-ph]} \BibitemShut {NoStop}%
\bibitem [{\citenamefont {Baek}\ \emph {et~al.}(2012)\citenamefont {Baek},
  \citenamefont {Ko},\ and\ \citenamefont {Park}}]{Baek:2011aa}%
  \BibitemOpen
  \bibfield  {author} {\bibinfo {author} {\bibfnamefont {S.}~\bibnamefont
  {Baek}}, \bibinfo {author} {\bibfnamefont {P.}~\bibnamefont {Ko}}, \ and\
  \bibinfo {author} {\bibfnamefont {W.-I.}\ \bibnamefont {Park}},\ }\href
  {\doibase 10.1007/JHEP02(2012)047} {\bibfield  {journal} {\bibinfo  {journal}
  {JHEP}\ }\textbf {\bibinfo {volume} {02}},\ \bibinfo {pages} {047} (\bibinfo
  {year} {2012})},\ \Eprint {http://arxiv.org/abs/1112.1847} {arXiv:1112.1847
  [hep-ph]} \BibitemShut {NoStop}%
\bibitem [{\citenamefont {Buckley}\ \emph {et~al.}(2015)\citenamefont
  {Buckley}, \citenamefont {Feld},\ and\ \citenamefont
  {Goncalves}}]{Buckley:2014fba}%
  \BibitemOpen
  \bibfield  {author} {\bibinfo {author} {\bibfnamefont {M.~R.}\ \bibnamefont
  {Buckley}}, \bibinfo {author} {\bibfnamefont {D.}~\bibnamefont {Feld}}, \
  and\ \bibinfo {author} {\bibfnamefont {D.}~\bibnamefont {Goncalves}},\ }\href
  {\doibase 10.1103/PhysRevD.91.015017} {\bibfield  {journal} {\bibinfo
  {journal} {Phys. Rev.}\ }\textbf {\bibinfo {volume} {D91}},\ \bibinfo {pages}
  {015017} (\bibinfo {year} {2015})},\ \Eprint {http://arxiv.org/abs/1410.6497}
  {arXiv:1410.6497 [hep-ph]} \BibitemShut {NoStop}%
\bibitem [{\citenamefont {Buchmueller}\ \emph {et~al.}(2015)\citenamefont
  {Buchmueller}, \citenamefont {Malik}, \citenamefont {McCabe},\ and\
  \citenamefont {Penning}}]{Buchmueller:2015eea}%
  \BibitemOpen
  \bibfield  {author} {\bibinfo {author} {\bibfnamefont {O.}~\bibnamefont
  {Buchmueller}}, \bibinfo {author} {\bibfnamefont {S.~A.}\ \bibnamefont
  {Malik}}, \bibinfo {author} {\bibfnamefont {C.}~\bibnamefont {McCabe}}, \
  and\ \bibinfo {author} {\bibfnamefont {B.}~\bibnamefont {Penning}},\ }\href
  {\doibase 10.1103/PhysRevLett.115.181802} {\bibfield  {journal} {\bibinfo
  {journal} {Phys. Rev. Lett.}\ }\textbf {\bibinfo {volume} {115}},\ \bibinfo
  {pages} {181802} (\bibinfo {year} {2015})},\ \Eprint
  {http://arxiv.org/abs/1505.07826} {arXiv:1505.07826 [hep-ph]} \BibitemShut
  {NoStop}%
\bibitem [{\citenamefont {Abdallah}\ \emph {et~al.}(2015)\citenamefont
  {Abdallah} \emph {et~al.}}]{Abdallah:2015ter}%
  \BibitemOpen
  \bibfield  {author} {\bibinfo {author} {\bibfnamefont {J.}~\bibnamefont
  {Abdallah}} \emph {et~al.},\ }\href {\doibase 10.1016/j.dark.2015.08.001}
  {\bibfield  {journal} {\bibinfo  {journal} {Phys. Dark Univ.}\ }\textbf
  {\bibinfo {volume} {9-10}},\ \bibinfo {pages} {8} (\bibinfo {year} {2015})},\
  \Eprint {http://arxiv.org/abs/1506.03116} {arXiv:1506.03116 [hep-ph]}
  \BibitemShut {NoStop}%
\bibitem [{\citenamefont {Abercrombie}\ \emph {et~al.}(2015)\citenamefont
  {Abercrombie} \emph {et~al.}}]{Abercrombie:2015wmb}%
  \BibitemOpen
  \bibfield  {author} {\bibinfo {author} {\bibfnamefont {D.}~\bibnamefont
  {Abercrombie}} \emph {et~al.},\ }\href@noop {} {\  (\bibinfo {year}
  {2015})},\ \Eprint {http://arxiv.org/abs/1507.00966} {arXiv:1507.00966
  [hep-ex]} \BibitemShut {NoStop}%
\bibitem [{\citenamefont {Busoni}\ \emph {et~al.}(2016)\citenamefont {Busoni}
  \emph {et~al.}}]{Boveia:2016mrp}%
  \BibitemOpen
  \bibfield  {author} {\bibinfo {author} {\bibfnamefont {G.}~\bibnamefont
  {Busoni}} \emph {et~al.},\ }\href@noop {} {\  (\bibinfo {year} {2016})},\
  \Eprint {http://arxiv.org/abs/1603.04156} {arXiv:1603.04156 [hep-ex]}
  \BibitemShut {NoStop}%
\bibitem [{\citenamefont {Englert}\ \emph {et~al.}(2016)\citenamefont
  {Englert}, \citenamefont {McCullough},\ and\ \citenamefont
  {Spannowsky}}]{Englert:2016joy}%
  \BibitemOpen
  \bibfield  {author} {\bibinfo {author} {\bibfnamefont {C.}~\bibnamefont
  {Englert}}, \bibinfo {author} {\bibfnamefont {M.}~\bibnamefont {McCullough}},
  \ and\ \bibinfo {author} {\bibfnamefont {M.}~\bibnamefont {Spannowsky}},\
  }\href {\doibase 10.1016/j.dark.2016.09.002} {\bibfield  {journal} {\bibinfo
  {journal} {Phys. Dark Univ.}\ }\textbf {\bibinfo {volume} {14}},\ \bibinfo
  {pages} {48} (\bibinfo {year} {2016})},\ \Eprint
  {http://arxiv.org/abs/1604.07975} {arXiv:1604.07975 [hep-ph]} \BibitemShut
  {NoStop}%
\bibitem [{\citenamefont {Albert}\ \emph {et~al.}(2017)\citenamefont {Albert}
  \emph {et~al.}}]{Bauer:2016gys}%
  \BibitemOpen
  \bibfield  {author} {\bibinfo {author} {\bibfnamefont {A.}~\bibnamefont
  {Albert}} \emph {et~al.},\ }\href {\doibase 10.1016/j.dark.2017.02.002}
  {\bibfield  {journal} {\bibinfo  {journal} {Phys. Dark Univ.}\ }\textbf
  {\bibinfo {volume} {16}},\ \bibinfo {pages} {49} (\bibinfo {year} {2017})},\
  \Eprint {http://arxiv.org/abs/1607.06680} {arXiv:1607.06680 [hep-ex]}
  \BibitemShut {NoStop}%
\bibitem [{\citenamefont {Baek}\ \emph {et~al.}(2017)\citenamefont {Baek},
  \citenamefont {Ko},\ and\ \citenamefont {Li}}]{Baek:2017vzd}%
  \BibitemOpen
  \bibfield  {author} {\bibinfo {author} {\bibfnamefont {S.}~\bibnamefont
  {Baek}}, \bibinfo {author} {\bibfnamefont {P.}~\bibnamefont {Ko}}, \ and\
  \bibinfo {author} {\bibfnamefont {J.}~\bibnamefont {Li}},\ }\href {\doibase
  10.1103/PhysRevD.95.075011} {\bibfield  {journal} {\bibinfo  {journal} {Phys.
  Rev.}\ }\textbf {\bibinfo {volume} {D95}},\ \bibinfo {pages} {075011}
  (\bibinfo {year} {2017})},\ \Eprint {http://arxiv.org/abs/1701.04131}
  {arXiv:1701.04131 [hep-ph]} \BibitemShut {NoStop}%
\bibitem [{\citenamefont {Alanne}\ and\ \citenamefont
  {Goertz}(2017)}]{Alanne:2017oqj}%
  \BibitemOpen
  \bibfield  {author} {\bibinfo {author} {\bibfnamefont {T.}~\bibnamefont
  {Alanne}}\ and\ \bibinfo {author} {\bibfnamefont {F.}~\bibnamefont
  {Goertz}},\ }\href@noop {} {\  (\bibinfo {year} {2017})},\ \Eprint
  {http://arxiv.org/abs/1712.07626} {arXiv:1712.07626 [hep-ph]} \BibitemShut
  {NoStop}%
\bibitem [{\citenamefont {Alitti}\ \emph {et~al.}(1991)\citenamefont {Alitti}
  \emph {et~al.}}]{Alitti:1990kw}%
  \BibitemOpen
  \bibfield  {author} {\bibinfo {author} {\bibfnamefont {J.}~\bibnamefont
  {Alitti}} \emph {et~al.} (\bibinfo {collaboration} {UA2}),\ }\href {\doibase
  10.1007/BF01570793} {\bibfield  {journal} {\bibinfo  {journal} {Z. Phys.}\
  }\textbf {\bibinfo {volume} {C49}},\ \bibinfo {pages} {17} (\bibinfo {year}
  {1991})}\BibitemShut {NoStop}%
\bibitem [{\citenamefont {Sirunyan}\ \emph {et~al.}(2017)\citenamefont
  {Sirunyan} \emph {et~al.}}]{Sirunyan:2017nvi}%
  \BibitemOpen
  \bibfield  {author} {\bibinfo {author} {\bibfnamefont {A.~M.}\ \bibnamefont
  {Sirunyan}} \emph {et~al.} (\bibinfo {collaboration} {CMS}),\ }\href@noop {}
  {\  (\bibinfo {year} {2017})},\ \Eprint {http://arxiv.org/abs/1710.00159}
  {arXiv:1710.00159 [hep-ex]} \BibitemShut {NoStop}%
\bibitem [{\citenamefont {Abe}\ \emph {et~al.}(1997)\citenamefont {Abe} \emph
  {et~al.}}]{Abe:1997hm}%
  \BibitemOpen
  \bibfield  {author} {\bibinfo {author} {\bibfnamefont {F.}~\bibnamefont
  {Abe}} \emph {et~al.} (\bibinfo {collaboration} {CDF}),\ }\href {\doibase
  10.1103/PhysRevD.55.R5263} {\bibfield  {journal} {\bibinfo  {journal} {Phys.
  Rev.}\ }\textbf {\bibinfo {volume} {D55}},\ \bibinfo {pages} {R5263}
  (\bibinfo {year} {1997})},\ \Eprint {http://arxiv.org/abs/hep-ex/9702004}
  {arXiv:hep-ex/9702004 [hep-ex]} \BibitemShut {NoStop}%
\bibitem [{\citenamefont {Aaltonen}\ \emph {et~al.}(2009)\citenamefont
  {Aaltonen} \emph {et~al.}}]{Aaltonen:2008dn}%
  \BibitemOpen
  \bibfield  {author} {\bibinfo {author} {\bibfnamefont {T.}~\bibnamefont
  {Aaltonen}} \emph {et~al.} (\bibinfo {collaboration} {CDF}),\ }\href
  {\doibase 10.1103/PhysRevD.79.112002} {\bibfield  {journal} {\bibinfo
  {journal} {Phys. Rev.}\ }\textbf {\bibinfo {volume} {D79}},\ \bibinfo {pages}
  {112002} (\bibinfo {year} {2009})},\ \Eprint {http://arxiv.org/abs/0812.4036}
  {arXiv:0812.4036 [hep-ex]} \BibitemShut {NoStop}%
\bibitem [{\citenamefont {Dobrescu}\ and\ \citenamefont
  {Yu}(2013)}]{Dobrescu:2013coa}%
  \BibitemOpen
  \bibfield  {author} {\bibinfo {author} {\bibfnamefont {B.~A.}\ \bibnamefont
  {Dobrescu}}\ and\ \bibinfo {author} {\bibfnamefont {F.}~\bibnamefont {Yu}},\
  }\href {\doibase 10.1103/PhysRevD.88.035021, 10.1103/PhysRevD.90.079901}
  {\bibfield  {journal} {\bibinfo  {journal} {Phys. Rev.}\ }\textbf {\bibinfo
  {volume} {D88}},\ \bibinfo {pages} {035021} (\bibinfo {year} {2013})},\
  \bibinfo {note} {[Erratum: Phys. Rev.D90,no.7,079901(2014)]},\ \Eprint
  {http://arxiv.org/abs/1306.2629} {arXiv:1306.2629 [hep-ph]} \BibitemShut
  {NoStop}%
\bibitem [{\citenamefont {Aad}\ \emph {et~al.}(2015)\citenamefont {Aad} \emph
  {et~al.}}]{Aad:2014aqa}%
  \BibitemOpen
  \bibfield  {author} {\bibinfo {author} {\bibfnamefont {G.}~\bibnamefont
  {Aad}} \emph {et~al.} (\bibinfo {collaboration} {ATLAS}),\ }\href {\doibase
  10.1103/PhysRevD.91.052007} {\bibfield  {journal} {\bibinfo  {journal} {Phys.
  Rev.}\ }\textbf {\bibinfo {volume} {D91}},\ \bibinfo {pages} {052007}
  (\bibinfo {year} {2015})},\ \Eprint {http://arxiv.org/abs/1407.1376}
  {arXiv:1407.1376 [hep-ex]} \BibitemShut {NoStop}%
\bibitem [{\citenamefont {Khachatryan}\ \emph {et~al.}(2016)\citenamefont
  {Khachatryan} \emph {et~al.}}]{Khachatryan:2016ecr}%
  \BibitemOpen
  \bibfield  {author} {\bibinfo {author} {\bibfnamefont {V.}~\bibnamefont
  {Khachatryan}} \emph {et~al.} (\bibinfo {collaboration} {CMS}),\ }\href
  {\doibase 10.1103/PhysRevLett.117.031802} {\bibfield  {journal} {\bibinfo
  {journal} {Phys. Rev. Lett.}\ }\textbf {\bibinfo {volume} {117}},\ \bibinfo
  {pages} {031802} (\bibinfo {year} {2016})},\ \Eprint
  {http://arxiv.org/abs/1604.08907} {arXiv:1604.08907 [hep-ex]} \BibitemShut
  {NoStop}%
\bibitem [{\citenamefont {Ackermann}\ \emph {et~al.}(2015)\citenamefont
  {Ackermann} \emph {et~al.}}]{Ackermann:2015zua}%
  \BibitemOpen
  \bibfield  {author} {\bibinfo {author} {\bibfnamefont {M.}~\bibnamefont
  {Ackermann}} \emph {et~al.} (\bibinfo {collaboration} {Fermi-LAT}),\ }\href
  {\doibase 10.1103/PhysRevLett.115.231301} {\bibfield  {journal} {\bibinfo
  {journal} {Phys. Rev. Lett.}\ }\textbf {\bibinfo {volume} {115}},\ \bibinfo
  {pages} {231301} (\bibinfo {year} {2015})},\ \Eprint
  {http://arxiv.org/abs/1503.02641} {arXiv:1503.02641 [astro-ph.HE]}
  \BibitemShut {NoStop}%
\bibitem [{\citenamefont {Alwall}\ \emph {et~al.}(2014)\citenamefont {Alwall},
  \citenamefont {Frederix}, \citenamefont {Frixione}, \citenamefont {Hirschi},
  \citenamefont {Maltoni}, \citenamefont {Mattelaer}, \citenamefont {Shao},
  \citenamefont {Stelzer}, \citenamefont {Torrielli},\ and\ \citenamefont
  {Zaro}}]{Alwall:2014hca}%
  \BibitemOpen
  \bibfield  {author} {\bibinfo {author} {\bibfnamefont {J.}~\bibnamefont
  {Alwall}}, \bibinfo {author} {\bibfnamefont {R.}~\bibnamefont {Frederix}},
  \bibinfo {author} {\bibfnamefont {S.}~\bibnamefont {Frixione}}, \bibinfo
  {author} {\bibfnamefont {V.}~\bibnamefont {Hirschi}}, \bibinfo {author}
  {\bibfnamefont {F.}~\bibnamefont {Maltoni}}, \bibinfo {author} {\bibfnamefont
  {O.}~\bibnamefont {Mattelaer}}, \bibinfo {author} {\bibfnamefont {H.~S.}\
  \bibnamefont {Shao}}, \bibinfo {author} {\bibfnamefont {T.}~\bibnamefont
  {Stelzer}}, \bibinfo {author} {\bibfnamefont {P.}~\bibnamefont {Torrielli}},
  \ and\ \bibinfo {author} {\bibfnamefont {M.}~\bibnamefont {Zaro}},\ }\href
  {\doibase 10.1007/JHEP07(2014)079} {\bibfield  {journal} {\bibinfo  {journal}
  {JHEP}\ }\textbf {\bibinfo {volume} {07}},\ \bibinfo {pages} {079} (\bibinfo
  {year} {2014})},\ \Eprint {http://arxiv.org/abs/1405.0301} {arXiv:1405.0301
  [hep-ph]} \BibitemShut {NoStop}%
\bibitem [{\citenamefont {Berlin}\ \emph {et~al.}(2014)\citenamefont {Berlin},
  \citenamefont {Hooper},\ and\ \citenamefont {McDermott}}]{Berlin:2014tja}%
  \BibitemOpen
  \bibfield  {author} {\bibinfo {author} {\bibfnamefont {A.}~\bibnamefont
  {Berlin}}, \bibinfo {author} {\bibfnamefont {D.}~\bibnamefont {Hooper}}, \
  and\ \bibinfo {author} {\bibfnamefont {S.~D.}\ \bibnamefont {McDermott}},\
  }\href {\doibase 10.1103/PhysRevD.89.115022} {\bibfield  {journal} {\bibinfo
  {journal} {Phys. Rev.}\ }\textbf {\bibinfo {volume} {D89}},\ \bibinfo {pages}
  {115022} (\bibinfo {year} {2014})},\ \Eprint {http://arxiv.org/abs/1404.0022}
  {arXiv:1404.0022 [hep-ph]} \BibitemShut {NoStop}%
\bibitem [{\citenamefont {Aprile}\ \emph {et~al.}(2017)\citenamefont {Aprile}
  \emph {et~al.}}]{Aprile:2017iyp}%
  \BibitemOpen
  \bibfield  {author} {\bibinfo {author} {\bibfnamefont {E.}~\bibnamefont
  {Aprile}} \emph {et~al.} (\bibinfo {collaboration} {XENON}),\ }\href
  {\doibase 10.1103/PhysRevLett.119.181301} {\bibfield  {journal} {\bibinfo
  {journal} {Phys. Rev. Lett.}\ }\textbf {\bibinfo {volume} {119}},\ \bibinfo
  {pages} {181301} (\bibinfo {year} {2017})},\ \Eprint
  {http://arxiv.org/abs/1705.06655} {arXiv:1705.06655 [astro-ph.CO]}
  \BibitemShut {NoStop}%
\bibitem [{\citenamefont {Petricca}\ \emph {et~al.}()\citenamefont {Petricca}
  \emph {et~al.}}]{Petricca:2017zdp}%
  \BibitemOpen
  \bibfield  {author} {\bibinfo {author} {\bibfnamefont {F.}~\bibnamefont
  {Petricca}} \emph {et~al.} (\bibinfo {collaboration} {CRESST}),\ }in\
  \href@noop {} {\emph {\bibinfo {booktitle} {{15th International Conference on
  Topics in Astroparticle and Underground Physics (TAUP 2017) Sudbury, Ontario,
  Canada, July 24-28, 2017}}}},\ \Eprint {http://arxiv.org/abs/1711.07692}
  {arXiv:1711.07692 [astro-ph.CO]} \BibitemShut {NoStop}%
\bibitem [{\citenamefont {Roberts}(2010)}]{Roberts:2010cj}%
  \BibitemOpen
  \bibfield  {author} {\bibinfo {author} {\bibfnamefont {B.~L.}\ \bibnamefont
  {Roberts}},\ }\bibfield  {booktitle} {\emph {\bibinfo {booktitle}
  {{Proceedings, 6th International Workshop on e+e- Collisions from Phi to Psi
  (PHIPSI09): Beijing, China, October 13-16, 2009}}},\ }\href {\doibase
  10.1088/1674-1137/34/6/021} {\bibfield  {journal} {\bibinfo  {journal} {Chin.
  Phys.}\ }\textbf {\bibinfo {volume} {C34}},\ \bibinfo {pages} {741} (\bibinfo
  {year} {2010})},\ \Eprint {http://arxiv.org/abs/1001.2898} {arXiv:1001.2898
  [hep-ex]} \BibitemShut {NoStop}%
\bibitem [{\citenamefont {Hagiwara}\ \emph {et~al.}(2011)\citenamefont
  {Hagiwara}, \citenamefont {Liao}, \citenamefont {Martin}, \citenamefont
  {Nomura},\ and\ \citenamefont {Teubner}}]{Hagiwara:2011af}%
  \BibitemOpen
  \bibfield  {author} {\bibinfo {author} {\bibfnamefont {K.}~\bibnamefont
  {Hagiwara}}, \bibinfo {author} {\bibfnamefont {R.}~\bibnamefont {Liao}},
  \bibinfo {author} {\bibfnamefont {A.~D.}\ \bibnamefont {Martin}}, \bibinfo
  {author} {\bibfnamefont {D.}~\bibnamefont {Nomura}}, \ and\ \bibinfo {author}
  {\bibfnamefont {T.}~\bibnamefont {Teubner}},\ }\href {\doibase
  10.1088/0954-3899/38/8/085003} {\bibfield  {journal} {\bibinfo  {journal} {J.
  Phys.}\ }\textbf {\bibinfo {volume} {G38}},\ \bibinfo {pages} {085003}
  (\bibinfo {year} {2011})},\ \Eprint {http://arxiv.org/abs/1105.3149}
  {arXiv:1105.3149 [hep-ph]} \BibitemShut {NoStop}%
\bibitem [{\citenamefont {Chen}\ \emph {et~al.}(2017)\citenamefont {Chen},
  \citenamefont {Pospelov},\ and\ \citenamefont {Zhong}}]{Chen:2017awl}%
  \BibitemOpen
  \bibfield  {author} {\bibinfo {author} {\bibfnamefont {C.-Y.}\ \bibnamefont
  {Chen}}, \bibinfo {author} {\bibfnamefont {M.}~\bibnamefont {Pospelov}}, \
  and\ \bibinfo {author} {\bibfnamefont {Y.-M.}\ \bibnamefont {Zhong}},\ }\href
  {\doibase 10.1103/PhysRevD.95.115005} {\bibfield  {journal} {\bibinfo
  {journal} {Phys. Rev.}\ }\textbf {\bibinfo {volume} {D95}},\ \bibinfo {pages}
  {115005} (\bibinfo {year} {2017})},\ \Eprint
  {http://arxiv.org/abs/1701.07437} {arXiv:1701.07437 [hep-ph]} \BibitemShut
  {NoStop}%
\bibitem [{\citenamefont {Leveille}(1978)}]{Leveille:1977rc}%
  \BibitemOpen
  \bibfield  {author} {\bibinfo {author} {\bibfnamefont {J.~P.}\ \bibnamefont
  {Leveille}},\ }\href {\doibase 10.1016/0550-3213(78)90051-2} {\bibfield
  {journal} {\bibinfo  {journal} {Nucl. Phys.}\ }\textbf {\bibinfo {volume}
  {B137}},\ \bibinfo {pages} {63} (\bibinfo {year} {1978})}\BibitemShut
  {NoStop}%
\bibitem [{\citenamefont {Kinoshita}\ and\ \citenamefont
  {Marciano}(1990)}]{Kinoshita:1990aj}%
  \BibitemOpen
  \bibfield  {author} {\bibinfo {author} {\bibfnamefont {T.}~\bibnamefont
  {Kinoshita}}\ and\ \bibinfo {author} {\bibfnamefont {W.~J.}\ \bibnamefont
  {Marciano}},\ }\href {\doibase 10.1142/9789814503273_0010} {\bibfield
  {journal} {\bibinfo  {journal} {Adv. Ser. Direct. High Energy Phys.}\
  }\textbf {\bibinfo {volume} {7}},\ \bibinfo {pages} {419} (\bibinfo {year}
  {1990})}\BibitemShut {NoStop}%
\bibitem [{\citenamefont {del Aguila}\ \emph {et~al.}(2008)\citenamefont {del
  Aguila}, \citenamefont {de~Blas},\ and\ \citenamefont
  {Perez-Victoria}}]{delAguila:2008pw}%
  \BibitemOpen
  \bibfield  {author} {\bibinfo {author} {\bibfnamefont {F.}~\bibnamefont {del
  Aguila}}, \bibinfo {author} {\bibfnamefont {J.}~\bibnamefont {de~Blas}}, \
  and\ \bibinfo {author} {\bibfnamefont {M.}~\bibnamefont {Perez-Victoria}},\
  }\href {\doibase 10.1103/PhysRevD.78.013010} {\bibfield  {journal} {\bibinfo
  {journal} {Phys. Rev.}\ }\textbf {\bibinfo {volume} {D78}},\ \bibinfo {pages}
  {013010} (\bibinfo {year} {2008})},\ \Eprint {http://arxiv.org/abs/0803.4008}
  {arXiv:0803.4008 [hep-ph]} \BibitemShut {NoStop}%
\bibitem [{\citenamefont {Freitas}\ \emph {et~al.}(2014)\citenamefont
  {Freitas}, \citenamefont {Lykken}, \citenamefont {Kell},\ and\ \citenamefont
  {Westhoff}}]{Freitas:2014pua}%
  \BibitemOpen
  \bibfield  {author} {\bibinfo {author} {\bibfnamefont {A.}~\bibnamefont
  {Freitas}}, \bibinfo {author} {\bibfnamefont {J.}~\bibnamefont {Lykken}},
  \bibinfo {author} {\bibfnamefont {S.}~\bibnamefont {Kell}}, \ and\ \bibinfo
  {author} {\bibfnamefont {S.}~\bibnamefont {Westhoff}},\ }\href {\doibase
  10.1007/JHEP09(2014)155, 10.1007/JHEP05(2014)145} {\bibfield  {journal}
  {\bibinfo  {journal} {JHEP}\ }\textbf {\bibinfo {volume} {05}},\ \bibinfo
  {pages} {145} (\bibinfo {year} {2014})},\ \bibinfo {note} {[Erratum:
  JHEP09,155(2014)]},\ \Eprint {http://arxiv.org/abs/1402.7065}
  {arXiv:1402.7065 [hep-ph]} \BibitemShut {NoStop}%
\bibitem [{\citenamefont {Patrignani}\ \emph {et~al.}(2016)\citenamefont
  {Patrignani} \emph {et~al.}}]{Patrignani:2016xqp}%
  \BibitemOpen
  \bibfield  {author} {\bibinfo {author} {\bibfnamefont {C.}~\bibnamefont
  {Patrignani}} \emph {et~al.} (\bibinfo {collaboration} {Particle Data
  Group}),\ }\href {\doibase 10.1088/1674-1137/40/10/100001} {\bibfield
  {journal} {\bibinfo  {journal} {Chin. Phys.}\ }\textbf {\bibinfo {volume}
  {C40}},\ \bibinfo {pages} {100001} (\bibinfo {year} {2016})}\BibitemShut
  {NoStop}%
\bibitem [{\citenamefont {Döbrich}\ \emph {et~al.}(2016)\citenamefont
  {Döbrich}, \citenamefont {Jaeckel}, \citenamefont {Kahlhoefer}, \citenamefont
  {Ringwald},\ and\ \citenamefont {Schmidt-Hoberg}}]{Dobrich:2015jyk}%
  \BibitemOpen
  \bibfield  {author} {\bibinfo {author} {\bibfnamefont {B.}~\bibnamefont
  {Döbrich}}, \bibinfo {author} {\bibfnamefont {J.}~\bibnamefont {Jaeckel}},
  \bibinfo {author} {\bibfnamefont {F.}~\bibnamefont {Kahlhoefer}}, \bibinfo
  {author} {\bibfnamefont {A.}~\bibnamefont {Ringwald}}, \ and\ \bibinfo
  {author} {\bibfnamefont {K.}~\bibnamefont {Schmidt-Hoberg}},\ }\href
  {\doibase 10.1007/JHEP02(2016)018} {\bibfield  {journal} {\bibinfo  {journal}
  {JHEP}\ }\textbf {\bibinfo {volume} {02}},\ \bibinfo {pages} {018} (\bibinfo
  {year} {2016})},\ \bibinfo {note} {[JHEP02,018(2016)]},\ \Eprint
  {http://arxiv.org/abs/1512.03069} {arXiv:1512.03069 [hep-ph]} \BibitemShut
  {NoStop}%
\bibitem [{\citenamefont {Dolan}\ \emph {et~al.}(2017)\citenamefont {Dolan},
  \citenamefont {Ferber}, \citenamefont {Hearty}, \citenamefont {Kahlhoefer},\
  and\ \citenamefont {Schmidt-Hoberg}}]{Dolan:2017osp}%
  \BibitemOpen
  \bibfield  {author} {\bibinfo {author} {\bibfnamefont {M.~J.}\ \bibnamefont
  {Dolan}}, \bibinfo {author} {\bibfnamefont {T.}~\bibnamefont {Ferber}},
  \bibinfo {author} {\bibfnamefont {C.}~\bibnamefont {Hearty}}, \bibinfo
  {author} {\bibfnamefont {F.}~\bibnamefont {Kahlhoefer}}, \ and\ \bibinfo
  {author} {\bibfnamefont {K.}~\bibnamefont {Schmidt-Hoberg}},\ }\href@noop {}
  {\  (\bibinfo {year} {2017})},\ \Eprint {http://arxiv.org/abs/1709.00009}
  {arXiv:1709.00009 [hep-ph]} \BibitemShut {NoStop}%
\bibitem [{\citenamefont {Bjorken}\ \emph {et~al.}(1988)\citenamefont
  {Bjorken}, \citenamefont {Ecklund}, \citenamefont {Nelson}, \citenamefont
  {Abashian}, \citenamefont {Church}, \citenamefont {Lu}, \citenamefont {Mo},
  \citenamefont {Nunamaker},\ and\ \citenamefont {Rassmann}}]{Bjorken:1988as}%
  \BibitemOpen
  \bibfield  {author} {\bibinfo {author} {\bibfnamefont {J.~D.}\ \bibnamefont
  {Bjorken}}, \bibinfo {author} {\bibfnamefont {S.}~\bibnamefont {Ecklund}},
  \bibinfo {author} {\bibfnamefont {W.~R.}\ \bibnamefont {Nelson}}, \bibinfo
  {author} {\bibfnamefont {A.}~\bibnamefont {Abashian}}, \bibinfo {author}
  {\bibfnamefont {C.}~\bibnamefont {Church}}, \bibinfo {author} {\bibfnamefont
  {B.}~\bibnamefont {Lu}}, \bibinfo {author} {\bibfnamefont {L.~W.}\
  \bibnamefont {Mo}}, \bibinfo {author} {\bibfnamefont {T.~A.}\ \bibnamefont
  {Nunamaker}}, \ and\ \bibinfo {author} {\bibfnamefont {P.}~\bibnamefont
  {Rassmann}},\ }\href {\doibase 10.1103/PhysRevD.38.3375} {\bibfield
  {journal} {\bibinfo  {journal} {Phys. Rev.}\ }\textbf {\bibinfo {volume}
  {D38}},\ \bibinfo {pages} {3375} (\bibinfo {year} {1988})}\BibitemShut
  {NoStop}%
\bibitem [{\citenamefont {Alekhin}\ \emph {et~al.}(2016)\citenamefont {Alekhin}
  \emph {et~al.}}]{Alekhin:2015byh}%
  \BibitemOpen
  \bibfield  {author} {\bibinfo {author} {\bibfnamefont {S.}~\bibnamefont
  {Alekhin}} \emph {et~al.},\ }\href {\doibase 10.1088/0034-4885/79/12/124201}
  {\bibfield  {journal} {\bibinfo  {journal} {Rept. Prog. Phys.}\ }\textbf
  {\bibinfo {volume} {79}},\ \bibinfo {pages} {124201} (\bibinfo {year}
  {2016})},\ \Eprint {http://arxiv.org/abs/1504.04855} {arXiv:1504.04855
  [hep-ph]} \BibitemShut {NoStop}%
\bibitem [{\citenamefont {Feng}\ \emph
  {et~al.}(2017{\natexlab{a}})\citenamefont {Feng}, \citenamefont {Galon},
  \citenamefont {Kling},\ and\ \citenamefont {Trojanowski}}]{Feng:2017uoz}%
  \BibitemOpen
  \bibfield  {author} {\bibinfo {author} {\bibfnamefont {J.}~\bibnamefont
  {Feng}}, \bibinfo {author} {\bibfnamefont {I.}~\bibnamefont {Galon}},
  \bibinfo {author} {\bibfnamefont {F.}~\bibnamefont {Kling}}, \ and\ \bibinfo
  {author} {\bibfnamefont {S.}~\bibnamefont {Trojanowski}},\ }\href@noop {} {\
  (\bibinfo {year} {2017}{\natexlab{a}})},\ \Eprint
  {http://arxiv.org/abs/1708.09389} {arXiv:1708.09389 [hep-ph]} \BibitemShut
  {NoStop}%
\bibitem [{\citenamefont {Feng}\ \emph
  {et~al.}(2017{\natexlab{b}})\citenamefont {Feng}, \citenamefont {Galon},
  \citenamefont {Kling},\ and\ \citenamefont {Trojanowski}}]{Feng:2017vli}%
  \BibitemOpen
  \bibfield  {author} {\bibinfo {author} {\bibfnamefont {J.~L.}\ \bibnamefont
  {Feng}}, \bibinfo {author} {\bibfnamefont {I.}~\bibnamefont {Galon}},
  \bibinfo {author} {\bibfnamefont {F.}~\bibnamefont {Kling}}, \ and\ \bibinfo
  {author} {\bibfnamefont {S.}~\bibnamefont {Trojanowski}},\ }\href@noop {} {\
  (\bibinfo {year} {2017}{\natexlab{b}})},\ \Eprint
  {http://arxiv.org/abs/1710.09387} {arXiv:1710.09387 [hep-ph]} \BibitemShut
  {NoStop}%
\bibitem [{\citenamefont {Pierog}\ \emph {et~al.}(2015)\citenamefont {Pierog},
  \citenamefont {Karpenko}, \citenamefont {Katzy}, \citenamefont {Yatsenko},\
  and\ \citenamefont {Werner}}]{Pierog:2013ria}%
  \BibitemOpen
  \bibfield  {author} {\bibinfo {author} {\bibfnamefont {T.}~\bibnamefont
  {Pierog}}, \bibinfo {author} {\bibfnamefont {I.}~\bibnamefont {Karpenko}},
  \bibinfo {author} {\bibfnamefont {J.~M.}\ \bibnamefont {Katzy}}, \bibinfo
  {author} {\bibfnamefont {E.}~\bibnamefont {Yatsenko}}, \ and\ \bibinfo
  {author} {\bibfnamefont {K.}~\bibnamefont {Werner}},\ }\href {\doibase
  10.1103/PhysRevC.92.034906} {\bibfield  {journal} {\bibinfo  {journal} {Phys.
  Rev.}\ }\textbf {\bibinfo {volume} {C92}},\ \bibinfo {pages} {034906}
  (\bibinfo {year} {2015})},\ \Eprint {http://arxiv.org/abs/1306.0121}
  {arXiv:1306.0121 [hep-ph]} \BibitemShut {NoStop}%
\bibitem [{\citenamefont {Baus}\ \emph {et~al.}()\citenamefont {Baus},
  \citenamefont {Pierog},\ and\ \citenamefont {Ulrich}}]{crmc}%
  \BibitemOpen
  \bibfield  {author} {\bibinfo {author} {\bibfnamefont {C.}~\bibnamefont
  {Baus}}, \bibinfo {author} {\bibfnamefont {T.}~\bibnamefont {Pierog}}, \ and\
  \bibinfo {author} {\bibfnamefont {R.}~\bibnamefont {Ulrich}},\ }\href@noop {}
  {\enquote {\bibinfo {title} {Cosmic ray monte carlo package},}\ }\bibinfo
  {howpublished} {\url{https://web.ikp.kit.edu/rulrich/crmc.html}}\BibitemShut
  {NoStop}%
\bibitem [{\citenamefont {Abbon}\ \emph {et~al.}(2007)\citenamefont {Abbon}
  \emph {et~al.}}]{Abbon:2007pq}%
  \BibitemOpen
  \bibfield  {author} {\bibinfo {author} {\bibfnamefont {P.}~\bibnamefont
  {Abbon}} \emph {et~al.} (\bibinfo {collaboration} {COMPASS}),\ }\href
  {\doibase 10.1016/j.nima.2007.03.026} {\bibfield  {journal} {\bibinfo
  {journal} {Nucl. Instrum. Meth.}\ }\textbf {\bibinfo {volume} {A577}},\
  \bibinfo {pages} {455} (\bibinfo {year} {2007})},\ \Eprint
  {http://arxiv.org/abs/hep-ex/0703049} {arXiv:hep-ex/0703049 [hep-ex]}
  \BibitemShut {NoStop}%
\bibitem [{\citenamefont {Essig}\ \emph {et~al.}(2010)\citenamefont {Essig},
  \citenamefont {Harnik}, \citenamefont {Kaplan},\ and\ \citenamefont
  {Toro}}]{Essig:2010gu}%
  \BibitemOpen
  \bibfield  {author} {\bibinfo {author} {\bibfnamefont {R.}~\bibnamefont
  {Essig}}, \bibinfo {author} {\bibfnamefont {R.}~\bibnamefont {Harnik}},
  \bibinfo {author} {\bibfnamefont {J.}~\bibnamefont {Kaplan}}, \ and\ \bibinfo
  {author} {\bibfnamefont {N.}~\bibnamefont {Toro}},\ }\href {\doibase
  10.1103/PhysRevD.82.113008} {\bibfield  {journal} {\bibinfo  {journal} {Phys.
  Rev.}\ }\textbf {\bibinfo {volume} {D82}},\ \bibinfo {pages} {113008}
  (\bibinfo {year} {2010})},\ \Eprint {http://arxiv.org/abs/1008.0636}
  {arXiv:1008.0636 [hep-ph]} \BibitemShut {NoStop}%
\bibitem [{\citenamefont {Lees}\ \emph {et~al.}(2016)\citenamefont {Lees} \emph
  {et~al.}}]{TheBABAR:2016rlg}%
  \BibitemOpen
  \bibfield  {author} {\bibinfo {author} {\bibfnamefont {J.~P.}\ \bibnamefont
  {Lees}} \emph {et~al.} (\bibinfo {collaboration} {BaBar}),\ }\href {\doibase
  10.1103/PhysRevD.94.011102} {\bibfield  {journal} {\bibinfo  {journal} {Phys.
  Rev.}\ }\textbf {\bibinfo {volume} {D94}},\ \bibinfo {pages} {011102}
  (\bibinfo {year} {2016})},\ \Eprint {http://arxiv.org/abs/1606.03501}
  {arXiv:1606.03501 [hep-ex]} \BibitemShut {NoStop}%
\bibitem [{\citenamefont {Aad}\ \emph {et~al.}(2014)\citenamefont {Aad} \emph
  {et~al.}}]{Aad:2014wra}%
  \BibitemOpen
  \bibfield  {author} {\bibinfo {author} {\bibfnamefont {G.}~\bibnamefont
  {Aad}} \emph {et~al.} (\bibinfo {collaboration} {ATLAS}),\ }\href {\doibase
  10.1103/PhysRevLett.112.231806} {\bibfield  {journal} {\bibinfo  {journal}
  {Phys. Rev. Lett.}\ }\textbf {\bibinfo {volume} {112}},\ \bibinfo {pages}
  {231806} (\bibinfo {year} {2014})},\ \Eprint {http://arxiv.org/abs/1403.5657}
  {arXiv:1403.5657 [hep-ex]} \BibitemShut {NoStop}%
\bibitem [{\citenamefont {Sirunyan}\ \emph {et~al.}(2018)\citenamefont
  {Sirunyan} \emph {et~al.}}]{Sirunyan:2018nnz}%
  \BibitemOpen
  \bibfield  {author} {\bibinfo {author} {\bibfnamefont {A.~M.}\ \bibnamefont
  {Sirunyan}} \emph {et~al.} (\bibinfo {collaboration} {CMS}),\ }\href@noop {}
  {\bibfield  {journal} {\bibinfo  {journal} {Submitted to: Phys. Lett.}\ }
  (\bibinfo {year} {2018})},\ \Eprint {http://arxiv.org/abs/1808.03684}
  {arXiv:1808.03684 [hep-ex]} \BibitemShut {NoStop}%
\bibitem [{\citenamefont {Abe}\ \emph {et~al.}(2010)\citenamefont {Abe} \emph
  {et~al.}}]{Abe:2010gxa}%
  \BibitemOpen
  \bibfield  {author} {\bibinfo {author} {\bibfnamefont {T.}~\bibnamefont
  {Abe}} \emph {et~al.} (\bibinfo {collaboration} {Belle-II}),\ }\href@noop {}
  {\  (\bibinfo {year} {2010})},\ \Eprint {http://arxiv.org/abs/1011.0352}
  {arXiv:1011.0352 [physics.ins-det]} \BibitemShut {NoStop}%
\bibitem [{Note5()}]{Note5}%
  \BibitemOpen
  \bibinfo {note} {One may also consider a naturalness criterion for $y_\chi $,
  but it is generally relaxed with respect to the $g_S^{\mu \mu }$ naturalness
  bound by a factor $(m_{\protect \rm DM}/M)^2$, where $m_{\protect \rm DM}$ is
  either $m_\chi $ or some UV scale in the dark sector (which may be
  significantly smaller than $M$).}\BibitemShut {Stop}%
\end{thebibliography}
